\documentclass[aps,twocolumn,showpacs,floatfix, superscriptaddress, reprint, footinbib, a4paper]{revtex4-2}

\usepackage{subfigure}
\usepackage{psfrag,graphicx}
\usepackage{pict2e}
\usepackage{dcolumn}
\usepackage{amsmath,amssymb,braket}
\usepackage{bm}
\usepackage{color}
\usepackage{latexsym}
\usepackage{epstopdf}
\usepackage{color}
\usepackage{comment}
\usepackage[english]{babel}
\UseRawInputEncoding
\usepackage{amsfonts}
\usepackage{bm}
\usepackage{natbib}
\usepackage{bbold}

\usepackage{enumerate}

\definecolor{myblue}{rgb}{0.2,0.2,0.8}
\definecolor{myzard}{cmyk}{0,0,0.05,0}
\definecolor{mywhite}{rgb}{1,1,1}
\definecolor{mywhite}{rgb}{1,1,1}
\definecolor{myred}{rgb}{1,0.,0.3}
\usepackage[colorlinks=true,citecolor=myblue,linkcolor=myred]{hyperref}

\definecolor{darkgreen}{rgb}{0.0, 0.4, 0.26}

\definecolor{mygrey}{gray}{0.35}
\definecolor{myblue}{rgb}{0.2,0.2,0.8}
\definecolor{myzard}{cmyk}{0,0,0.05,0}
\definecolor{mywhite}{rgb}{1,1,1}
\definecolor{mywhite}{rgb}{1,1,1}
\definecolor{myred}{rgb}{1,0.,0.3}

\def\be{\begin{equation}}
\def\ee{\end{equation}}
\def\ba{\begin{align}}
\def\enda{\end{align}}
\def\bi{\begin{itemize}}
\def\ei{\end{itemize}}

\def\beq{\begin{equation}}
\def\beq{\begin{equation}}
\def\eeq{\end{equation}}

\DeclareMathOperator{\tr}{Tr}

\def\ii{{\bm i}}
\def\jj{{\bm j}}

\def\kk{{\textbf k}}

\newcommand{\spin}[2]{{\sigma}_{\bm{#1}}^{#2}}

\begin{document}

\title{Estimating ground-state properties in quantum simulators with global control}

 \author{Cristian Tabares}
 \email{cristian.tabares@csic.es}
 \affiliation{Instituto de F\'isica Fundamental IFF-CSIC, C. Serrano 113b, 28006 Madrid, Spain}
\affiliation{Max-Planck-Institut f\"{u}r Quantenoptik, Hans-Kopfermann-Str. 1, D-85748 Garching, Germany}
\author{Dominik S. Wild}
 \affiliation{Max-Planck-Institut f\"{u}r Quantenoptik, Hans-Kopfermann-Str. 1, D-85748 Garching, Germany}
 \affiliation{Munich Center for Quantum Science and Technology (MCQST), Schellingstr. 4, D-80799 Munich, Germany}
\author{J. Ignacio Cirac}
 \affiliation{Max-Planck-Institut f\"{u}r Quantenoptik, Hans-Kopfermann-Str. 1, D-85748 Garching, Germany}
 \affiliation{Munich Center for Quantum Science and Technology (MCQST), Schellingstr. 4, D-80799 Munich, Germany}
\author{Peter~Zoller}
 \affiliation{Institute for Theoretical Physics, University of Innsbruck, 6020 Innsbruck, Austria}
 \affiliation{Institute for Quantum Optics and Quantum Information of the Austrian Academy of Sciences, 6020 Innsbruck, Austria}
 \author{Alejandro Gonz\'alez-Tudela}
 \email{a.gonzalez.tudela@csic.es}
 \affiliation{Instituto de F\'isica Fundamental IFF-CSIC, C. Serrano 113b, 28006 Madrid, Spain}
 
\author{Daniel~Gonz\'alez-Cuadra}
\email{daniel.gonzalez@ift.csic.es}
\affiliation{Instituto de F\'isica Te\'orica UAM-CSIC, C. Nicol\'as Cabrera 13-15, Cantoblanco, 28049 Madrid, Spain}
\affiliation{Department of Physics, Harvard University, Cambridge, MA 02138, USA}

\begin{abstract}
Accurately determining ground-state properties of quantum many-body systems remains one of the major challenges of quantum simulation. In this work, we present a protocol for estimating the ground-state energy using only global time evolution under a target Hamiltonian. This avoids the need for controlled operations that are typically required in conventional quantum phase estimation and extends the algorithm applicability to analog simulators. Our method extracts energy differences from measurements of the Loschmidt echo over an initial ground-state approximation, combines them with direct energy measurements, and solves a set of equations to infer the individual eigenenergies. We benchmark this protocol on free-fermion systems, showing orders-of-magnitude precision gains over direct energy measurements on the initial state, with accuracy improving rapidly with initial-state fidelity and persisting for hundreds of modes. We further demonstrate applicability to the 2D Ising and Fermi-Hubbard models and show that the approach extends naturally to other observables such as order parameters. Finally, we analyze the effect of experimental imperfections and propose error-mitigation strategies.  These results establish a practical route to compute physically relevant quantities with high precision using globally controlled quantum simulators.
\end{abstract}

\maketitle

\paragraph*{Introduction.---}
Quantum simulation is one of the most promising applications of quantum computers~\cite{Altman_2021, Daley_2022}, providing access to quantum many-body regimes relevant to condensed-matter physics~\cite{Bauer_2020}, quantum chemistry~\cite{McArdle_2020}, and particle physics~\cite{DiMeglio_2024} that lie beyond the reach of classical computation. In particular, analog quantum simulators enable the efficient emulation of the dynamics of specific model Hamiltonians, whose ground states can be prepared either by cooling the system or through adiabatic protocols. Using these techniques, a wide range of quantum many-body phenomena have been experimentally explored using ultracold atoms in optical lattices~\cite{Gross_2017}, Rydberg atom arrays~\cite{Browaeys_2020}, trapped ions~\cite{Monroe_2021}, and superconducting qubits~\cite{Kjaergaard_2020}.

While analog devices are generally more robust to errors than their digital counterparts~\cite{Flannigan_2022, Trivedi_2024, Schiffer_2024, Kashyap_2025}, achieving high-fidelity state preparation remains challenging due to finite coherence times, experimental imperfections, and, in particular, the absence of quantum error correction. Since computing ground-state properties with high precision is essential for addressing open problems in many-body physics, developing methods to accurately estimate them from imperfectly prepared states is of great importance. In digital quantum computers, this task can be accomplished through quantum phase estimation~\cite{Kitaev_1995}, which requires long, ancilla-controlled evolutions. Although several variants have been proposed to reduce quantum resource requirements by shifting part of the workload to classical post-processing, these approaches cannot be directly implemented in globally controlled quantum simulators~\cite{OBrien_2019, Ge_2019, Somma_2019,Lin_2022, Dutkiewicz_2022, Dong_2022, Blunt_2023, clinton2024quantumphaseestimationcontrolled,Maskara_2025, Schiffer_2025,scali2025spectralsubspaceextractionincoherent} 
or do not work for arbitrary states~\cite{RobledoMoreno2025SQD,yu2025quantumcentricalgorithmsamplebasedkrylov, cavallar2025phasesensitivemeasurementsfermihubbardquantum}.

\begin{figure}[tb]
    \centering
    \includegraphics[width=\linewidth]{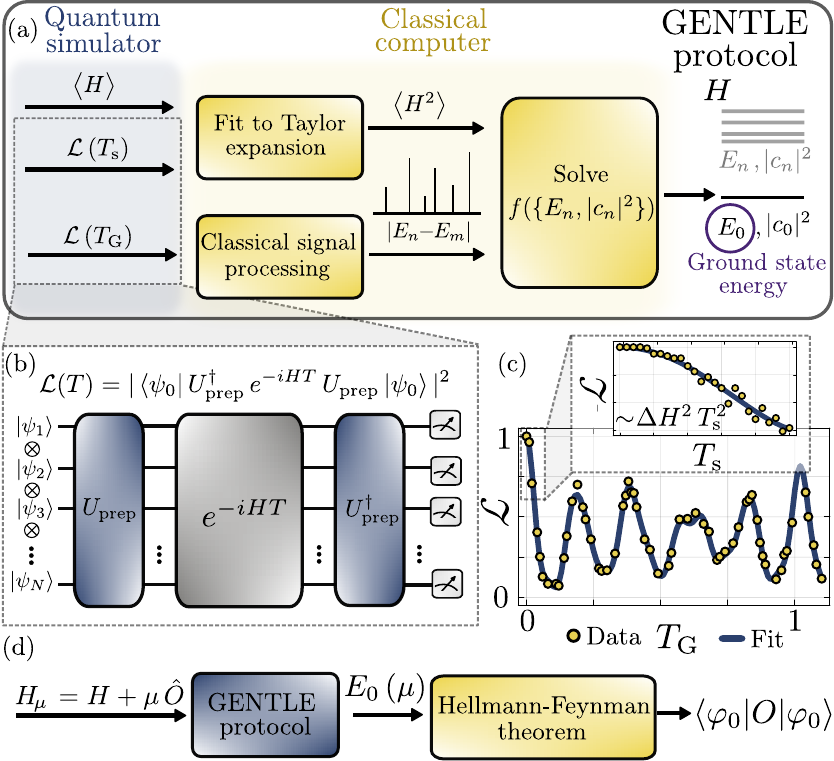}
    \caption{ 
    (a) Scheme of the protocol. Using the quantum simulator, the energy of the target Hamiltonian $\langle H \rangle$ is measured on an approximate ground state $\ket{\psi}$, together with the Loschmidt echo $\mathcal{L}(T)$ up to two different evolution times, $T_\text{s}$ and $T_\text{G}$, with $T_\text{s} \ll T_\text{G}$. From the first two quantities, $\langle H^2 \rangle$ is estimated through a Taylor expansion. A signal-processing analysis of $\mathcal{L}(T_\text{G})$ yields the associated spectrum, from which the energy differences between the eigenstates of $H$ are obtained. Together with $\langle H \rangle$ and $\langle H^2 \rangle$, these provide a system of equations $f(\{E_n,|c_n|^2\})=0$ whose solution allows the estimation of individual energies. (b) Quantum circuit illustrating the measurement of $\mathcal{L}(T)$. (c) Classical post-processing involves fitting the measured signals $\mathcal{L}(T_\text{G})$ and $\mathcal{L}(T_\text{s})$ and applying a short-time expansion to extract $\langle H^2 \rangle$ from the latter. (d) Applying the protocol to the perturbed Hamiltonian $H_\mu = {H} + \mu O$ provides access to the ground-state energy as a function of $\mu$, $E_0(\mu)$. The ground-state expectation value of $O$ is obtained from it using the Hellmann-Feynman theorem.
    }
    \label{fig:1}
\end{figure}

In this work, we address these challenges by introducing a ground-state energy estimation protocol that requires only global time evolution under the target Hamiltonian. Similar to previous approaches, our method relies on evolving an initial approximation to the ground state and measuring the corresponding Loschmidt echo [see Fig.~\ref{fig:1}(a)], which has been demonstrated experimentally in different analog devices~\cite{Jurcevic_2017, Singh_2019, Guo_2021, Braumuller_2022, Cao_2022, Zhu_2024,Roberts2024, Karch_2025}. However, rather than determining the associated phase to extract the ground-state energy, an approach that generally demands controlled evolutions, we instead use the echo to obtain energy differences via classical post-processing of the measured signal. The ground-state energy is then inferred by solving a nonlinear system of equations, combining the echo data with an additional measurement of the energy on the initial state. Furthermore, we show how our protocol can be easily extended to estimate observables beyond the energy, such as order parameters of interest in condensed-matter systems.

We benchmark our protocol on free fermions, showing that, once the initial state fidelity reaches a certain threshold, we improve the ground-state energy estimation by orders-of-magnitude compared with direct energy measurements. Furthermore, we show that this improvement persists as we increase the system size to hundreds of modes. We then apply our protocol to the 2D Ising model and doped Fermi-Hubbard ladders, illustrating its applicability to non-integrable and strongly correlated regimes. Finally, we analyze the impact of experimental imperfections, including depolarizing and shot noise, and show how their effects can be mitigated.

\paragraph*{Ground State Energy Through Loschmidt Echoes (GENTLE).---}
Similarly to other phase-estimation protocols, our approach begins with the preparation of an approximate ground state of the target Hamiltonian $H$, denoted by $\ket{\psi}$. Expanding this state in the eigenbasis of $H = \sum_n E_n \ket{\varphi_n}\bra{\varphi_n}$, we can write $\ket{\psi} = \sum_n c_n \ket{\varphi_n}$, with $\ket{\varphi_0}$ being the exact ground state of $H$. We assume that $\ket{\psi}$ can be prepared from a product state $\ket{\psi_0} = \bigotimes_i \ket{\psi_i}$ via a unitary operation $U_\text{prep}$, i.e., $\ket{\psi} = U_\text{prep} \ket{\psi_0}$. 

The protocol consists of two main steps [see Fig.~\ref{fig:1}(a)]. In the first step, the quantum simulator is used to measure certain quantities on the initially prepared state $\ket{\psi}$. In the second step, the resulting data is classically post-processed to estimate properties of the exact ground state $\ket{\varphi_0}$. During the first step, $\ket{\psi}$ is evolved for a time $T_\text{G}$ under the target Hamiltonian $H$ to measure the corresponding Loschmidt echo, $\mathcal{L}(T_\text{G}) = |\langle \psi | e^{-i T_\text{G} H} | \psi \rangle|^2$, which can be efficiently measured by reversing the state preparation and projecting onto the initial product basis [Fig.~\ref{fig:1}(b)]. We provide further details on this procedure in the Supplementary Material (SM)~\cite{SupMatGENTLE}, including schemes to implement $U^\dagger_\text{prep}$ in analog quantum simulators as well as alternative measurement protocols. In the following, we consider for concreteness that $U_\text{prep}$ corresponds to an adiabatic evolution with total duration $T_{\mathrm{a}}$. 

Expanding $\ket{\psi}$ in the eigenbasis of $H$, the Loschmidt echo can be expressed as
\begin{equation}
\mathcal{L}(T_\text{G}) = \mathcal{L}_0 + \sum_{n< m} 2\, p_n p_m \cos{\left[(E_n - E_m) T_\text{G}\right]},
\end{equation}
where $p_n = |c_n|^2$. In the End Matter (EM), we describe in detail how both the energy differences $E_n - E_m$ and the amplitudes $p_n p_m$ can be extracted from the measured echo using standard signal processing techniques. Specifically, we employ a combination of compressed sensing~\cite{Candes2006a,Candes2006b,Candes2006c} and a nonlinear fit to first reconstruct the spectrum of the signal for a finite number of measured times and shots, and then relate the extracted frequencies to the energy differences by solving the associated minimum distance superset problem~\cite{Fontoura2018MDSP}. For this final step, we require that $p_0 > p_n$ for all $n > 0$.

The echo data is combined with additional measurements of $\langle H \rangle = \sum_n p_n E_n$ and $\langle {H}^2 \rangle = \sum_n p_n E_n^2$, forming a system of nonlinear equations, $f(\{E_n,|c_n|^2\})=0$, which we solve numerically to estimate the individual energies [see Fig.~\ref{fig:1}(a)]. While $\langle {H} \rangle$ can be directly measured on most devices, $\langle {H}^2 \rangle$ can be efficiently extracted from a short-time expansion of the echo [see Fig.~\ref{fig:1}(c)],
\begin{equation}
\mathcal{L}(T_\text{s}) = 1 - \left[\langle {H}^2 \rangle - \langle {H} \rangle^2\right] T_\text{s}^2 + \mathcal{O}(T_\text{s}^4)\,,
\end{equation}
as well as with classical shadows~\cite{Huang2020}.

In summary, our GENTLE protocol enables estimation of the ground-state energy by measuring the echo up to two different times, $T_\text{G}$ and $T_\text{s}$ with $T_\text{s} \ll T_\text{G}$, in combination with a measurement of $\langle H \rangle$ on the initially prepared state. All of these quantities can be directly accessed on a quantum device capable of implementing global time evolutions under $H$, making the protocol particularly well-suited to analog quantum simulators. 

In addition to the ground-state energy, our approach can be naturally extended to estimate the expectation value of other observables, such as order parameters, on the ground state, i.e., $\bra{\varphi_0}{O}\ket{\varphi_0}$. This can be achieved by first applying the GENTLE protocol to the perturbed Hamiltonian ${H}_\mu = {H} + \mu\,{O}$ to estimate the corresponding ground-state energy $E_0(\mu)$. The target observable is then obtained via a straightforward application of the Hellmann-Feynman theorem,  
\begin{equation}~\label{eq:HF_theorem}
     \langle \varphi_0 | {O} | \varphi_0 \rangle = \frac{\mathrm{d} E_\mu}{\mathrm{d} \mu} \bigg|_{\mu=0}.
\end{equation}
We note that this approach is specially efficient when the time evolution under ${O}$ can be easily implemented on the quantum simulator.

\begin{figure}[tb]
    \centering
    \includegraphics[width=\linewidth]{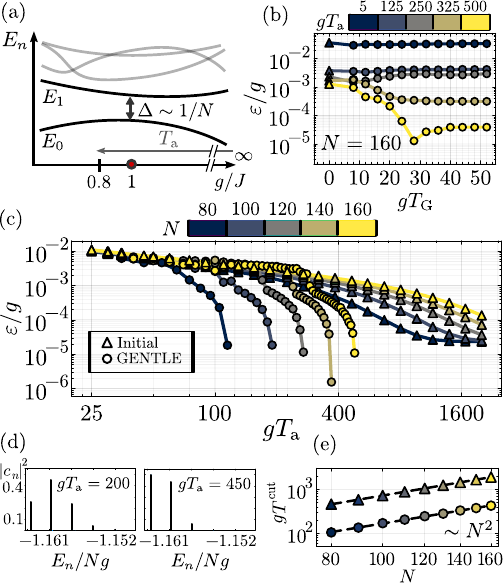}
    \caption{(a) Sketch of the energy spectrum of the 1D TFIM~\eqref{eq:H_Ising} as a function of $g/J$, showing the energy gap $\Delta = E_1 - E_0$ closing at the critical point (red dot). The arrow indicates the adiabatic ramp, of duration $gT_{\mathrm{a}}$, used to prepare the state $\ket{\psi}$ from a product state. (b) Residual energy $\varepsilon/g$ obtained using the GENTLE protocol applied to $\ket{\psi}$ as a function of $gT_{\mathrm{G}}$, for different preparation times $gT_{\mathrm{a}}$. We consider a chain of $N = 160$ spins and use echo time steps of $g\Delta T_{\mathrm{G}} = 0.1$. The values shown at $gT_{\mathrm{a}} = 0$ (triangles) correspond to the energy of the initial state. (c) Residual energy $\varepsilon/g$ before (initial state) and after applying the GENTLE protocol (with fixed $gT_{\mathrm{G}} = 24$) as a function of $gT_{\mathrm{a}}$ for different system sizes $N$. (d) Low-energy local density of states for two states prepared with different $gT_{\mathrm{a}}$ in a chain of $N = 160$ spins. (e) Total evolution time $gT^{\mathrm{cut}}$ required to reach an energy accuracy $|\tilde{E_0} - E_0|/g < 0.025$ as a function of $N$, comparing cases with and without the GENTLE protocol.}
    \label{fig:2}
\end{figure}

\paragraph*{Scaling of the protocol for an integrable model.---}
We now investigate how the performance of the protocol scales with different parameters, including the fidelity of the initial state, the GENTLE evolution time $T_G$, and the total system size $N$. To benchmark the protocol for large systems, we first apply it to the transverse-field Ising model (TFIM),
\begin{equation}~\label{eq:H_Ising}
    {H}_{\mathrm{TFIM}} = -J \sum_{\langle \bm{i}, \bm{j} \rangle} \spin{i}{x} \spin{j}{x} - g \sum_{\bm{i}} \spin{i}{z},
\end{equation}
where $\spin{i}{x}$ and $\spin{i}{z}$ are Pauli matrices acting on site $\bm{i}$. 

We consider in particular a 1D chain with open boundary conditions, where the model can be mapped to a free-fermion system that can be efficiently solved on a classical computer~\cite{Surace2002_free_fermions}. We initialize the system in the product state $\ket{\psi_0} = \ket{\uparrow}^{\otimes N}$, corresponding to the ground state of Eq.~\eqref{eq:H_Ising} for $J = 0$, and perform a linear adiabatic ramp to the point $g/J = 0.8$ over a total evolution time $gT_a$ [Fig.~\ref{fig:2}(a)]. Since this trajectory crosses the critical point at $g/J = 1.0$ separating the paramagnetic and the ferromagnetic phases, where the energy gap closes as $\sim 1/N$~\cite{Sachdev2011}, the fidelity of the final state $\ket{\psi}$ after the evolution depends on the ramp duration $gT_a$. 

We compare the ground-state energy estimation obtained using the GENTLE protocol applied to $\ket{\psi}$ with that obtained from a direct measurement of ${H}$ on $\ket{\psi}$. As a figure of merit, we use the residual energy, $\varepsilon = |\tilde{E}_0-E_0|/N$, where $\tilde{E}_0$ and $E_0$ are the estimated and the exact ground-state energies, respectively. Fig.~\ref{fig:2}(b) shows the residual energy $\varepsilon$ obtained after applying the GENTLE protocol to initial states prepared with different adiabatic times $gT_\text{a}$, as a function of the GENTLE evolution time $gT_\text{G}$ used to extract the echo. We observe that when $gT_\text{a}$ is sufficiently large, corresponding to an initial state with high overlap with the exact ground state, $\varepsilon/g$ improves rapidly with increasing $gT_\text{G}$ over several orders of magnitude, before eventually saturating (the origin of this saturation is discussed in the EM).

Fig.~\ref{fig:2}(c) shows the residual energy $\varepsilon/g$ as a function of the adiabatic preparation time $gT_\text{a}$ for a fixed GENTLE evolution time $gT_\text{G}$ and for different system sizes $N$. Two distinct regimes can be identified, consistent with Fig.~\ref{fig:2}(b): for small $gT_\text{a}$, applying the GENTLE protocol does not significantly improve the energy of the initial state; in contrast, once the initial overlap with the ground state becomes sufficiently large, $\varepsilon/g$ decreases much faster with $gT_\text{a}$ compared to the energy of the adiabatically prepared state. We further observe that this improvement, spanning several orders of magnitude in precision, is maintained even for systems with hundreds of sites, with the crossover point between the two regimes shifting to larger $gT_\text{a}$ as $N$ increases. To gain further insight into this change of scaling, Fig.~\ref{fig:2}(d) displays the local density of states $\delta_{\psi}(E) = \sum_n |c_n|^2 \delta(E-E_n)$ for two different preparation times $gT_\text{a}$, representative of each regime. We find that only in the second regime the ground state provides the dominant contribution to the initial state, thereby validating the main assumption underlying our method.

Finally, in Fig.~\ref{fig:2}(e) we analyze the scaling of the protocol with system size. We plot the minimum total evolution time $gT^{\mathrm{cut}} = g(2T_{\mathrm{a}} + T_{\mathrm{G}})$ required to reach an energy accuracy $|\tilde{E}_0 - E_0|/g < 0.025$, comparing the results of adiabatic state preparation and the GENTLE protocol~\footnote{Here we are comparing the total evolution time required for one repetition of GENTLE. While the protocol has to be repeated to achieve a certain precision, this is not a problem for current quantum devices.  The bottleneck lies in the simulated evolution time, limited by the experimental coherence time, which our protocol minimizes.}. In the adiabatic case this accuracy scales with the inverse of the gap, $1/\Delta^2 \sim N^2$. The GENTLE protocol exhibits the same asymptotic scaling, but with a substantially smaller prefactor. This demonstrates that, as long as the good-overlap condition is satisfied, the GENTLE protocol achieves the same energy accuracy with a significantly shorter total evolution time.

\begin{figure}[tb]
    \centering
    \includegraphics[width=\linewidth]{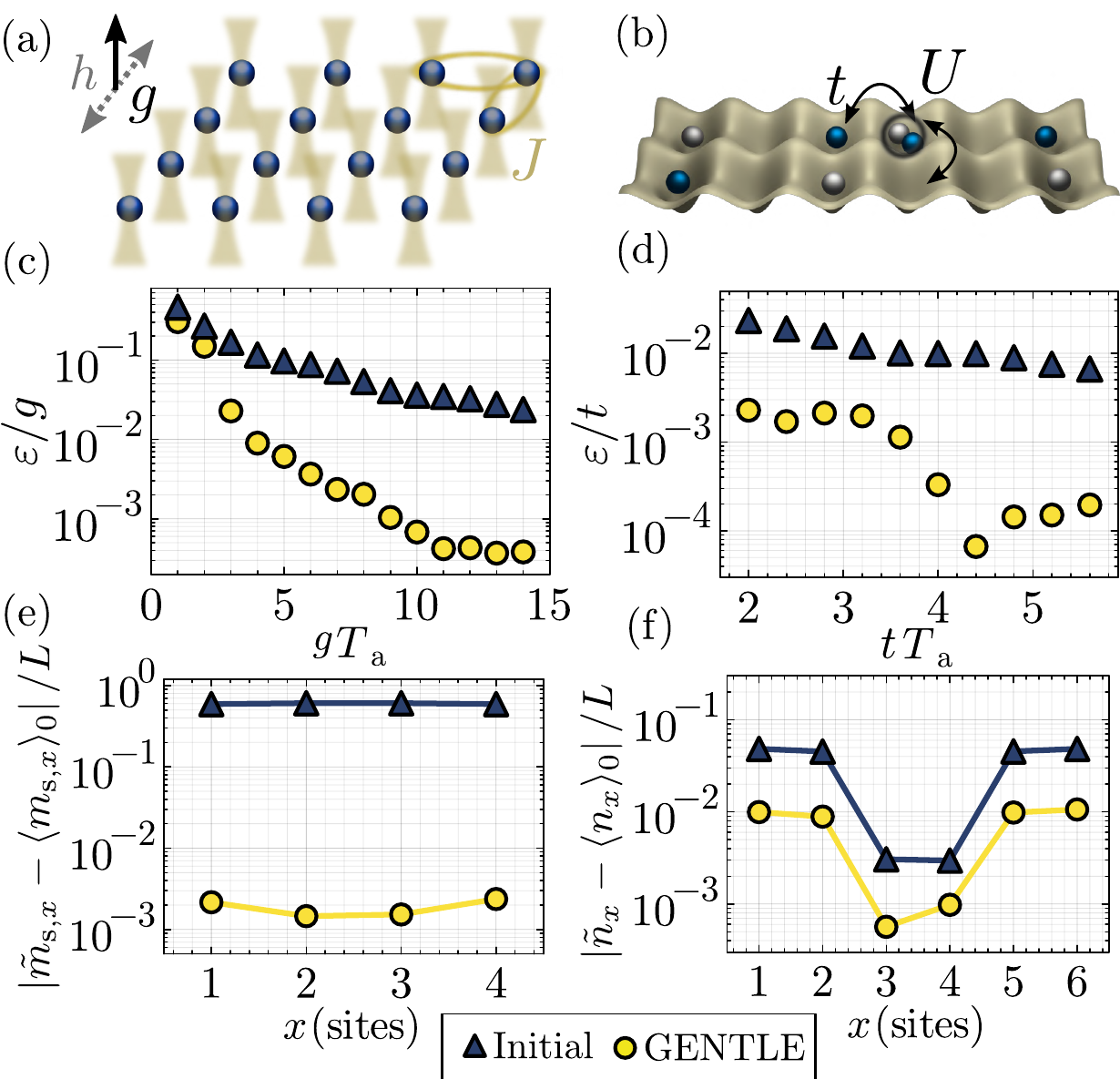}
    \caption{(a) The TFIM can be implemented using neutral atoms in optical tweezer arrays (red), whereas (b) the FH model is naturally realized with fermionic atoms in optical lattices (blue). (c) Residual energy $\varepsilon$ obtained for a $4\times4$ TFIM at $J/g=-1$ as a function of the initial preparation time $gT_\mathrm{a}$, comparing the energy measured on the initial state (triangles) and after applying the GENTLE protocol (circles). During the state preparation, a small staggered field ${H}_\mathrm{s}=h\,\sum_{x,y} (-1)^{x+y}{\sigma}^{z}_{(x,y)}$, with $h = 0.0025$, is added to lift the ground-state degeneracy. We compute the echo up to $gT_{\mathrm{G}}=40$. (d) Analogous results for a $2\times6$ FH model with eight electrons at $U/t=8$. We also compute the echo up to $t T_{\mathrm{G}}=40$.  (e) Error in the staggered magnetization ${m}_{\mathrm{s},x}$, averaged along $L$ rows in the $y$ direction, for the TFIM after a state preparation with $gT_\mathrm{a}=20$, illustrating the improvement achieved by applying the modified GENTLE protocol to ${H}(\mu)$ (see main text). (f) Same comparison for the FH ladder, showing the error in the average density along the $y$ direction for $tT_\mathrm{a}=5.6$.}
    \label{fig:3}
\end{figure}

\paragraph*{The protocol beyond integrable models.---}
After benchmarking the protocol in a free-fermion system, we now apply it to non-integrable models, and further use it to estimate observables beyond the ground-state energy. We consider two representative cases: an antiferromagnetic (AF) TFIM in a 2D square lattice [Eq.~\eqref{eq:H_Ising} with $J<0$], and the Fermi-Hubbard (FH) model on a ladder geometry, described by the Hamiltonian
\begin{equation*}
{H}_{\mathrm{FH}}=-t\sum_{\langle\ii,\jj\rangle,\sigma} {c}_{\ii,\sigma}^\dagger {c}^{\vphantom{\dagger}}_{\jj,\sigma}+U \sum_{\ii}{n}_{\ii,\uparrow}{n}_{\ii,\downarrow}-\sum_{\ii,\sigma}\mu_{\bm{i},\sigma}{n}_{\ii,\sigma}\label{eq:HFH}\,,
\end{equation*}
where $c^{(\dagger)}_{\ii,\sigma}$ denote the (creation) annihilation fermionic operators corresponding to the $\sigma\in\{\uparrow,\downarrow\}$ spin component at site $\ii$, and ${n}_{\ii,\sigma} = {c}_{\ii,\sigma}^\dagger {c}^{\vphantom{\dagger}}_{\ii,\sigma}$. While the TFIM can be implemented across various experimental platforms, such as Rydberg atom arrays~\cite{Browaeys_2020} [Fig.~\ref{fig:3}(a)], the FH model is naturally realized with ultracold atoms in optical lattices~\cite{Gross_2017} [Fig.~\ref{fig:3}(b)].

Figs.~\ref{fig:3}(c) and~(d) show the results obtained by applying the GENTLE protocol to estimate the ground-state energy of these two models. For the TFIM, the initial state is prepared adiabatically following a similar trajectory as in Fig.~\ref{fig:2}, but ending in the AF phase at $g/J = -1.0$. For the doped FH ladder at $U/t = 8$, inspired by the experimental procedure in Ref.~\cite{Xu2025_nature}, we start from a band-insulator state that is converted into a product of singlets on the occupied sites while maintaining a chemical potential $\mu/t=4$ on rungs $2$ and $5$~\cite{Tabares2025}. We then ramp the tunneling between dimers over a time $tT_t=4$ and subsequently lower the chemical potentials during a final time $tT_a$. As in the free-fermion case, the residual energy $\varepsilon$ obtained after applying GENTLE exhibits a much faster convergence with the initial fidelity compared to the energy measured on the initial state, yielding in both cases an improvement of more than two orders of magnitude.

Finally, we apply our protocol in combination with the Hellmann-Feynman theorem to estimate relevant order parameters. For the TFIM, we consider the staggered magnetization averaged along columns, ${m}_{\mathrm{s},x} = \sum_{y} (-1)^{x+y}{\sigma}^{x}_{(x,y)}$, which serves as the order parameter of the AF phase. For the FH ladder, we analyze the average occupation along rungs, ${n}_x = \sum_{y,\sigma}{n}_{(x,y),\sigma}$, where a charge-density-wave profile signals the emergence of stripe ordering. In both cases, these observables can be incorporated into the Hamiltonian during the evolution, ${H}(\mu) = {H} + \mu {O}$, as required by our protocol. The corresponding results are shown in Figs.~\ref{fig:3}(e) and~(f), demonstrating that the protocol significantly improves the estimation of the order parameters compared to direct measurements on the initial states. To obtain these results, we apply the protocol to the modified Hamiltonian with $\mu\in\{0, \pm0.1, \pm0.2\}$ and fit the results to a second order Taylor expansion, recovering the derivative and the corresponding order parameter~\eqref{eq:HF_theorem} as the linear coefficient.

\begin{figure}[tb]
    \centering
    \includegraphics[width=\linewidth]{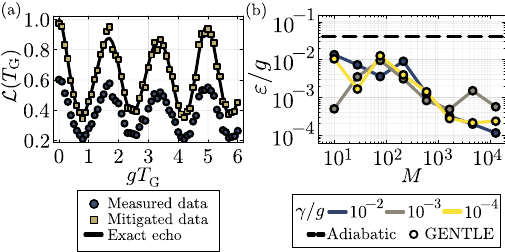}
    \caption{(a) Loschmidt echo as a function of $gT_{\mathrm{G}}$ for a $5\times5$ TFIM, starting from an initial state prepared via an adiabatic ramp of duration $gT_{\mathrm{a}} = 12$ ending at $J/g = 1$. A global depolarizing channel $\mathcal{M}_t$ with strength $\gamma/g = 10^{-2}$ is applied during both the state preparation and the GENTLE evolution, and $M = 600$ measurements are performed per data point. We display the measured noisy Loschmidt echo, the value recovered using the echo-verification technique introduced in the text, and the exact noiseless result. (b) Residual energy $\varepsilon$ as a function of $M$ in the Loschmidt echo measured up to $g\,T_{\mathrm{G}}=31$, for different values of the depolarizing strength $\gamma$, and after applying the echo-verification procedure.}
    \label{fig:4}
\end{figure}

\paragraph*{The effect of noise.---}
Here we analyze the impact of global depolarizing noise and shot noise on the performance of the protocol. Specifically, we examine how these noise sources affect the estimation of the ground-state energy of the 2D TFIM~\eqref{eq:H_Ising} at $J/g = 1$, using the same state-preparation scheme as in Fig.~\ref{fig:3}. Fig.~\ref{fig:4}(a) displays the measured Loschmidt echo in the presence of a depolarizing channel with $\gamma/g = 10^{-2}$, as detailed in the EM, and a measurement budget of $M = 600$ per data point. As expected, the echo starts from a value smaller than one due to the initial state depolarization and exhibits a gradual damping of its oscillations. After applying an echo-verification procedure to mitigate the noise~\cite{Yang2023,Yang2024PhaseMeas, Schiffer_2025}, the recovered signal closely matches the exact noiseless echo. This corrected echo can then be used to reliably execute the GENTLE protocol, as shown in Fig.~\ref{fig:4}(b), where we study the influence of different measurement budgets $M$ and depolarizing strengths $\gamma$. These results demonstrate that, given a sufficient number of measurements, the error-mitigation scheme successfully restores the Loschmidt echo largely independently of the noise strength. This is particularly relevant since current quantum hardware typically exhibits $10^{-3} < \gamma < 10^{-2}$, already allowing improvements of nearly two orders of magnitude in the estimated energy when moderate measurement budgets of $M \sim 500$-$1000$ shots are used. Finally, in the SM~\cite{SupMatGENTLE} we also analyze the effect of finite temperature initial states in the accuracy of the protocol, demonstrating that it still works under experimentally reasonable temperatures~\cite{Xu2025_nature}.

\paragraph*{Conclusions and outlook.---}
In this work, we introduced an algorithm to estimate ground-state properties that requires only the implementation of global time evolution under the target Hamiltonian on an approximate ground state, followed by measurements of the corresponding Loschmidt echo. By applying classical signal-processing techniques to the measured data, we achieve orders-of-magnitude improvements in estimating both the ground-state energy and relevant order parameters, as demonstrated for the 2D Ising and Fermi-Hubbard models. Looking ahead, it would be interesting to extend the protocol to estimate observables whose time evolution cannot be easily implemented, as well as to perform estimations at finite temperature. Achieving this without relying on ancillas or controlled evolutions would expand the capabilities of analog devices to perform high-precision simulations of quantum many-body systems.

\paragraph*{Acknowledgments.---} C.T. acknowledges useful discussions with Philipp Preiss, Benjamin Schiffer, and Rahul Trivedi during his time at MPQ, and also with Jan T. Schneider. D.G.-C. acknowledges financial support through the Ram\'on y Cajal Program (RYC2023-044201-I), financed by MICIU/AEI/10.13039/501100011033 and by the FSE+. C.T. acknowledges support from Comunidad de Madrid (PIPF-2022/TEC-25625) and also from Fundaci\'on Humanismo y Ciencia.  C.T. and A.G.T. acknowledge support from the the Spanish National Research Council (CSIC) Research Platform on Quantum Technologies PTI-001, and from Spanish project PID2021-127968NB-I00 funded by the Ministry of Science, Innovation and Universities (MICIU)-State Research Agency (AEI) 10.13039/501100011033, and from the QUANTERA project MOLAR with reference PCI2024-153449 and funded by MICIU–AEI 10.13039/501100011033 and by the EU. P.Z. acknowledges support by the EU Horizon Europe programs HORIZON-CL4-2022-QUANTUM-02-SGA via the project 101113690 (PASQuanS2.1). D.S.W. and J.I.C. acknowledge support from the German Federal Ministry of Education and Research (BMBF) through the funded project ALMANAQC, grant number 13N17236 within the research program “Quantum Systems”. We also acknowledge funding from the project FermiQP (13N15889). The work at MPQ is partly funded by THEQUCO as part of the Munich Quantum Valley, which is supported by the Bavarian state government with funds from the Hightech Agenda Bayern Plus. The numerical calculations in this paper use the ITensor~\cite{itensor} and the F\_utilities~\cite{Surace2002_free_fermions} packages and were executed on the FinisTerrae supercomputer supported by the Centro de Supercomputaci\'on de Galicia (CESGA).

\bibliographystyle{apsrev4-2}
\bibliography{references}

\clearpage
\section{END MATTER}
\paragraph*{The GENTLE protocol.---~\label{sec:appendixA}}
In this section, we describe in detail the practical implementation of the GENTLE protocol, which consists of four main steps: (1) data acquisition, (2) fitting of the Loschmidt echo, (3) mapping of the signal parameters to physical observables, and (4) solution of the GENTLE equations.

(1) In the first step, the unitary operator $U_{\mathrm{prep}}$ is used to prepare the initial state $\ket{\psi}$, an approximation to the ground state of the Hamiltonian ${H} = \sum_n E_n \ket{\varphi_n}\bra{\varphi_n}$, starting from a product state $\ket{\psi_0}$. The protocol then consists of evolving $\ket{\psi}$ under ${H}$ for a duration $T_\text{G}$ and measuring the corresponding Loschmidt echo,
\begin{align}~\label{eq:end_matter_echo} 
\mathcal{L}(T_\text{G}) &= |\langle \psi | e^{-i {H}T_\text{G}}\ket{\psi}\rangle|^2 \\ &= \mathcal{L}_0 + \sum_{n\neq m} 2 \,p_n\, p_m \cos{\left[\left(E_n - E_m\right)T_\text{G}\right]}\nonumber\,, 
\end{align}
where $p_n = |\braket{\psi|\varphi_n}|^2 = |c_n|^2$.
We are free to choose an arbitrary set of evolution times $t \in \mathcal{T} = \{ t_n \}$ (not necessarily equally spaced), and for each $t_n$ we obtain one data point of the signal. The measured values of the Loschmidt echo $\mathcal{L}(t_n)$ are denoted by $y_n$, with $n \in [N]$, so that the complete measured signal can be represented as a vector $y = [y_0, \dots, y_{N-1}]^T$. Without loss of generality, each entry can be decomposed as $y_n = y_n^0 + z_n$, where $z_n$ represents the measurement noise.

(2) The second step of the protocol exploits the fact that the Loschmidt echo $\mathcal{L}(t)$ belongs to a family of parametrized functions of the form
\begin{equation}~\label{eq:end_matter_ffun_GENTLE}
    f_{\bm{A},\bm{\omega}}(t) = A_0 + \sum_k A_k\,\cos{\left(\omega_k t\right)} \,,
\end{equation}
where $\bm{A} = (A_0, A_1, \dots)$ and $\bm{\omega} = (0, \omega_1, \dots)$ are the amplitude and frequency vectors, respectively. If the $t_n$ are evenly spaced, the noise-free part of the measured signal, $y^0$, can be written as 
\begin{equation}~\label{eq:end_matter_decomposition_signal}
    y_n^0 = \sum_{k\in \mathcal{K}} \beta_k\,a_k\, \cos{\left[\frac{\pi}{N}\left(n+\frac{1}{2}\right)k\right]}=\sum_{k\in\mathcal{K}}\beta_k a_k\,C_n(k)\,.
\end{equation}
Otherwise, the $n+1/2$ term is replaced with the appropiate $t_n$. In Eq.~\eqref{eq:end_matter_decomposition_signal}, $\mathcal{K}$ denotes the frequency support, and $\beta_k$ is a normalization factor defined as $\beta_0 = 1/\sqrt{N}$ for $k=0$ and $\beta_k = \sqrt{2/N}$ otherwise. The equation is general as long as $k \in \mathbb{R}$. For algorithmic purposes, however, we discretize $\mathcal{K}$ and define the orthogonal matrix $C_{kn} =\beta_k C_n(k)$ with $n,k \in [N]$. If all frequencies $k$ in Eq.~\eqref{eq:end_matter_decomposition_signal} belong to this discrete set, the signal is said to be on-grid; otherwise, it is off-grid. In the on-grid case, the signal can be expressed in the frequency domain as $x^0 = C^{\dagger}y^0$~(see SM \cite{SupMatGENTLE} for further details).

The goal of this second step is to accurately reconstruct $x^0$ using as few signal samples as possible. Even for coherent data, the measured signal $y = y^0 + z$ contains three types of noise: (i) shot noise, due to a finite number of measurements; (ii) off-grid noise, since the true frequencies generally do not coincide with the discretized ones; and (iii) truncation noise, arising from keeping fewer frequencies in $x^0$ than are actually present in the signal. Furthermore, data collection is often limited by the low repetition rates and the finite coherence times of current quantum hardware.

To address these challenges, we employ a compressed sensing framework~\cite{Candes2006a,Candes2006b,Candes2006c}, which formulates the recovery of the sparse signal as an optimization problem, and we further discuss its advantages in the SM~\cite{SupMatGENTLE}.  Compressed sensing uses signal sparsity to reconstruct $x^0$ from very few samples, provided the sampling times $\mathcal{T}$ are chosen randomly. Specifically, if $\lVert z\rVert_2 < \eta$ for $\eta \in \mathbb{R}$, the signal in the frequency domain can be efficiently recovered by solving the convex optimization problem
\begin{equation}~\label{eq:appendix_CS_program}
    \min_{x\in\mathbb{R}^N}\lVert x\rVert_1 \quad\mathrm{s.t.}\quad \lVert Cx - y\rVert_2\leq\eta\,,
\end{equation}
where $\lVert\cdot\rVert_p$ denotes the $p$-norm. The solution $x^*$ to this program satisfies $\lVert x^* - x^0\rVert_2 \leq A\cdot\eta$, with $A$ a constant factor. In practice, the noise bound $\eta$ is unknown. We initialize the solver with a lower-bound estimate (for $M$ measurements, this scales as $\mathcal{O}(1/\sqrt{M})$) and solve the quadratic program using a conic solver~\cite{Odonoghue2016,Odonoghue2021,Odonoghue2023}. If the constraint is not satisfied within a fixed number of iterations, $\eta$ is gradually increased until convergence

After obtaining a preliminary spectrum, we apply a first correction by merging frequencies that are very close (typically within $0.05$-$0.2$ in energy units, which are chosen such that $g=1$ in case of the TFIM and $t =1$ for the FH model) and discarding amplitudes below a chosen threshold ($\sim5\cdot10^{-3
}$). To account for residual off-grid effects due to discretization (taken with resolution $\mathcal{O}(1/\sqrt{M})$, lower-bounded by $0.05$ in energy units), we perform a final nonlinear least-squares fit of the signal to the model $f_{\bm{A},\bm{\omega}}$,~\cite{Ding2023QCELS,ding2024robustgroundstateenergyestimation}:
\begin{equation*}
\begin{split}
    \left(\bm{A^*},\bm{\omega^*}\right) = \mathrm{argmin}_{\bm{A}\in\mathbb{R}^P,\bm{\omega}\in\mathbb{R}^P} \left(\sum_{n=1}^{N} \left|y_n - f_{\bm{A},\bm{\omega}}(t_n)\right|^2\right)\,,
\end{split}
\end{equation*}
using the Levenberg-Marquardt algorithm~\cite{Gavin2013TheLM}. The amplitudes $\bm{A}$ and frequencies $\bm{\omega}$ obtained from compressed sensing serve as initial guesses, yielding a refined set $(\bm{A}^*, \bm{\omega}^*)$ that accurately reproduces the measured Loschmidt echo.

(3) The third step of the protocol connects the fitted signal parameters to the physical quantities of interest, namely the energy differences $|E_i - E_j|$ and the corresponding amplitudes $2 p_ip_j$. Specifically, given a set of $P$ frequencies $\bm{\omega}$ such that $\omega_p = |E_i - E_j|$, the goal is to determine the smallest set of energies ${E_k}$ whose pairwise differences reproduce all frequencies in $\bm{\omega}$. In computational mathematics, this task is known as the minimum distance superset problem~\cite{Fontoura2018MDSP}. 

In practice, we address a relaxed version of the problem by assuming that the lowest energy level $E_{k_0}$ contributes as a reference to all other energy differences; that is, for any $E_k \in \{E_k\}$ with $k \neq k_0$, we have $|E_{k_0} - E_k| \in \bm{\omega}$. The reconstruction is then performed using a greedy algorithm that explores all possible configurations consistent with this assumption, allowing a maximum frequency mismatch of $5\cdot10^{-3
}$ (in energy units). This condition corresponds to the situation where the overlap of the initial state with the ground state dominates over the others, i.e., $|c_{0}| > |c_n|$ for all eigenstates contributing to $\ket{\psi}$.

(4) The fourth and final step of the GENTLE protocol consists in solving the GENTLE equations: a combined set of $2P$ equations,
\begin{equation}
~\label{eq:end_matter_fitlosch}
    |E_i - E_j| = \omega_p^*\quad\mathrm{and}\quad
    2\,p_i\,p_j = A_p^*\,,
\end{equation}
together with
\begin{equation}~\label{eq:end_matter_Hs_GENTLE}
    \braket{{H}^k}=  p_0\,E^k_0+\sum_n p_n\,E^k_n\,,
\end{equation}
for $k = 1, 2$. Since experimental measurements are subject to statistical noise, these equations may not admit an exact solution. In practice, we determine an approximate one by minimizing the sum of the residues of Eqs.~\eqref{eq:end_matter_fitlosch} and~\eqref{eq:end_matter_Hs_GENTLE} using the L-BFGS optimization algorithm~\cite{NLopt,Liu1989}. To mitigate the risk of converging to local minima, we initialize the solver near the expected solution setting the initial guess for the ground-state energy $E_0$ equal to $\langle{H}\rangle$, while assigning the remaining energies as $E_0 + \omega_p$. The initial overlaps are chosen as $p_0 = 1/2$, with the remaining $p_n$ distributed so that $\sum_n p_n = 1$.

\begin{figure}[tb]
    \centering
    \includegraphics[width=\linewidth]{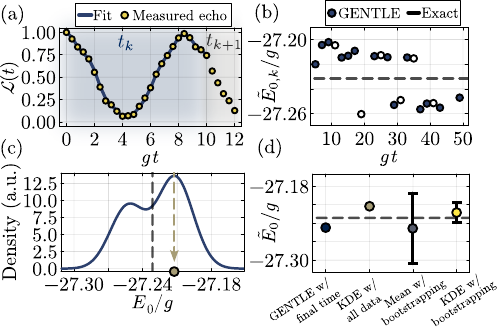}
    \caption{(a) Loschmidt echo for the 2D TFIM considered in Fig.~\ref{fig:3} of the main text, starting from an initial state prepared with $gT_a = 10$ and using $M = 10^3$ measurements per point. (b) Estimated energies $\tilde{E}_{0,k}$ obtained by applying the protocol to each time $t_k$. White points indicate values randomly discarded in a single bootstrap iteration. (c) Probability density obtained via KDE applied to all data in (b); the maximum, representing the most probable energy, is marked with a grey dot. (d) Final $\tilde{E}_0$ after successive postprocessing steps. Bootstrapping with 70\% of the data and $N_{\mathrm{boot}} = 10^4$ iterations is used to determine the error bars. The exact value of the energy is shown as a black dashed line.}
    \label{fig:end_matter}
\end{figure}

Finally, we note that the quantities obtained through this protocol may still carry some error due to local minima in the system solution. To minimize it, we divide the total Loschmidt echo, measured over a total evolution time $t$, into $K$ segments starting from $t_0$ with spacing $\Delta t$ (set throughout the paper as $t_0 = 5$ and $\Delta t = 1$, in energy units
) [Fig.~\ref{fig:end_matter}(a)]. We then apply the GENTLE protocol to each segment to obtain an array of ground-state energy estimations, $(E_{0,1}, E_{0,2}, \dots, E_{0,K})$ [Fig.~\ref{fig:end_matter}(b)]. Assuming these values are independent samples drawn from a common underlying probability distribution, we employ Kernel Density Estimation (KDE)~\cite{Gramacki2018KDE} to reconstruct its probability density. The location of the maximum, i.e., the most probable energy, is taken as the final energy estimate. The resolution of the numerical implementation used imposes a difference of $10^{-3}$ between KDE energies, leading to the saturation found in Fig.~\ref{fig:2}(c) of the main text. The resulting distribution obtained from all data in Fig.~\ref{fig:end_matter}(b) is shown in Fig.~\ref{fig:end_matter}(c).

To further enhance robustness, we apply a bootstrapping technique~\cite{davison1997bootstrap}: at each iteration, 70\% of the data are randomly resampled, and the above analysis is repeated $10^3$-$10^4$ times. The final reported energy corresponds to the mean of the resulting distribution, while the standard deviation provides the uncertainty. As shown in Fig.~\ref{fig:end_matter}(d), this procedure not only yields the most accurate energy estimates, but also produces realistic and well-calibrated error bars.

\paragraph*{Noise modeling and error mitigation.---~\label{sec:appendixB}}
To investigate the impact of experimental errors, we model the effect of a global depolarizing channel that transforms the unitary evolution $\mathcal{U}_t$ into a quantum channel $\mathcal{M}_t$:
\begin{equation}\label{eq:end_matter_M_t}
\mathcal{M}_t(\rho)=e^{-\gamma\, t}\,\mathcal{U}_t\,\rho\, \mathcal{U}^\dagger_t+\frac{\left(1-e^{-\gamma \,t}\right)}{2M} I_{2M}\,,
\end{equation}
where $\rho \in \mathbb{C}^{2M\times 2M}$ is the density matrix, $\gamma>0$ characterizes the noise strength, and $I_{2M}$ is the identity matrix. This channel affects both the state preparation and the echo evolution, leading to an exponential damping of the noiseless Loschmidt echo: $\mathcal{L}_{\mathcal{M}}(t) \approx e^{-2\,\gamma \,T_{\mathrm{a}}}\,e^{-\gamma \,t}\mathcal{L}(t)$.
In addition, each measured point of the Loschmidt echo is subject to shot noise, since only a finite number of measurements $M$ is performed. Each measurement yields a $0$ or $1$ with probability $\mathcal{L}(t)$, so the outcome is modeled as a binomial distribution with $M$ samples.

To recover the noiseless echo from the noisy data, we employ a quantum error mitigation technique known as echo verification~\cite{Cai2023_QEM,OBrien2021,OBrien2023_EV_exp,Yang2023,Yang2024PhaseMeas,ding2024robustgroundstateenergyestimation, Schiffer_2025}. The idea is to run the quantum circuit forward and backward in time and then project back onto the initial state. In the absence of noise, the final state perfectly reproduces the initial state, but hardware imperfections lead to an exponential decay of the survival probability. By fitting this decay to a function $g(t) = Ae^{-Bt}$, where $A$ and $B$ are fit parameters and $t$ is the total evolution time, one can estimate the effective depolarizing parameter. The measured noisy echo can then be corrected as $\mathcal{L}(T_G) \approx A^{-1}e^{+B t}\,\mathcal{L}_{\mathcal{M}}(T_{\mathrm{G}})$, providing an accurate approximation to the noiseless echo.

\clearpage
\appendix

\begin{widetext}

\begin{center}
\textbf{\large Supplementary Material: Estimating ground-state properties in quantum simulators with global control \\}
\end{center}
\setcounter{equation}{0}
\setcounter{figure}{0}
\makeatletter

\renewcommand{\thefigure}{SM\arabic{figure}}
\renewcommand{\thesection}{SM\arabic{section}}  
\renewcommand{\theequation}{SM\arabic{equation}}  
In this Supplementary Material, we provide a detailed analysis of several topics introduced in the main text. Section~\ref{secSM:StatePreparation} discusses how to measure overlaps in analog quantum simulators using only global control. Since this is a crucial step in our algorithm, necessary to measure Loschmidt echoes, we review existing proposals in the literature and present a more efficient protocol for implementing $U^\dagger_\text{prep}$. Section~\ref{secSM:CompressedSensingAndNumberOfMeasurements} explains our choice of the compressed sensing (CS) framework as the classical signal-processing tool for analyzing the echo. In particular, Sec.~\ref{subsec:intro_CS} provides an in-depth introduction to CS and highlights its expected sampling advantages for our problem, while Sec.~\ref{subsec:sampling_adv} explicitly demonstrates these advantages in a concrete example by benchmarking CS against alternative signal-processing techniques. Finally, Sec.~\ref{secSM:ExperimentalAnalysis} examines finite-temperature effects on the GENTLE protocol, which are relevant in various quantum simulation platforms. We demonstrate the robustness of the protocol by first deriving conditions for its validity with finite-temperature initial states (Sec.~\ref{subsec:SM_thermal_analytic}), and then numerically benchmarking these conditions in two representative examples (Sec.~\ref{subsec:SM_thermal_numeric}).

\section{Measuring state overlaps in globally controlled quantum simulators~\label{secSM:StatePreparation}}

\subsection{Protocols for the measurement of state overlaps}

A key assumption of the GENTLE protocol is the ability to measure state overlaps, that is, to project a state $\ket{\phi_1}$ onto another state $\ket{\phi_2}$ and measure the probability $|\braket{\phi_1|\phi_2}|^2$. The standard approaches to determine these quantities can be broadly grouped into three families: (1) protocols that create copies of the state \emph{in space}, (2) protocols that create copies of the state \emph{in time}, and (3) protocols based on randomized measurements.

\begin{enumerate}
\item The first family of protocols requires preparing the states $\ket{\phi_1}$ and $\ket{\phi_2}$ in two distinct, spatially separated parts of the system. These two parts are then coupled via a beamsplitter operation, and single-site measurements are used to reconstruct the overlap. In the context of ultracold atoms in optical lattices, this technique was originally proposed to measure Renyi entropies of many-body states~\cite{MouraAlves2004,Daley2012,Pichler_2013}, and it has been experimentally demonstrated~\cite{Islam_2015,Kaufman_2016,Lukin_2019,Su2023}. In digital quantum simulators, the same idea is interpreted as a destructive SWAP test implemented through Bell-pair measurements on each qubit pair~\cite{Cincio_2018,Subasi_2019,Huggins2021,boucomas2024measuringtemporalentanglementexperiments}, and has been successfully realized in superconducting circuits~\cite{OBrien2023_EV_exp}.

\item The second family of protocols employs a single quantum register to prepare both $\ket{\phi_1}$ and $\ket{\phi_2}$. The additional requirement is the ability to implement the unitaries $U_1$ and $U_2^\dagger$, defined through $\ket{\phi_i} = U_i \ket{\phi_0}$ for $i=1,2$, with $\ket{\phi_0}$ a product state. Since $\bra{\phi_i} = \bra{\phi_0} U_i^\dagger$, the overlap can be rewritten as $|\braket{\phi_1|\phi_2}|^2 = |\braket{\phi_0|U_1^\dagger\,U_2|\phi_0}|^2$, which corresponds to the probability of measuring the configuration associated with $\ket{\phi_0}$ after applying the unitary $U_1^\dagger\,U_2$ and projecting back to the original basis. The overlap can then be estimated as $|\braket{\phi_1|\phi_2}|^2 \approx M_{\phi_0}/M$, where $M$ is the total number of measurements and $M_{\phi_0}$ is the number of outcomes corresponding to the $\ket{\phi_0}$ configuration. This method,often referred to as the compute-uncompute technique in the digital quantum simulation literature~\cite{Gentinetta2024overheadconstrained,Yang_2024}, has been successfully implemented in superconducting circuits~\cite{Havlicek2019,Mi2021,Braumuller2022,abanin2025constructiveinterferenceedgequantum,zhang2025quantumcomputationmoleculargeometry} and trapped-ion platforms~\cite{Jurcevic_2017,Garttner2017,Gilmore2021,Hemery2024}.

\item Finally, there exists a distinct family of methods to estimate state overlaps based on randomized measurements~\cite{Rath2021,Elben2023,Naldesi2024_randomized}. These protocols operate by repeatedly preparing and measuring a quantum state in randomly chosen bases, and then reconstructing properties of the state from the resulting measurement statistics. They do not directly measure the overlap $|\braket{\phi_1|\phi_2}|^2$, but rather infer it from the statistical correlations in the data. Such protocols have been successfully implemented in recent experiments~\cite{Elben2020,Zhu2022,Joshi2023,Vitale2024,votto2025learningmixedquantumstates}, and the measurement of the Loschmidt echo represents a natural application of their framework.

\end{enumerate}

Given that current quantum simulators are typically constrained either by system size or by limited connectivity, the second family of protocols is, in principle, more suitable for measuring the Loschmidt echo. Provided that the unitary $U_{\mathrm{prep}}$ used to prepare an approximate ground state $\ket{\psi}$ of a Hamiltonian ${H}$ from a product state $\ket{\psi_0}$ (and its inverse $U_{\mathrm{prep}}^\dagger$) can be implemented, the Loschmidt echo over $\ket{\psi}$ can be expressed as
\begin{equation}~\label{eq:echo_SM_uprep}
\mathcal{L}(T_G) = |\left\langle\psi\right|e^{-i{H}T_G}\left|\psi\right\rangle|^2 = |\left\langle\psi_0\right|U_{\mathrm{prep}}^\dagger ,e^{-i{H}T_G},U_{\mathrm{prep}}\left|\psi_0\right\rangle|^2\,,
\end{equation}
which experimentally corresponds to applying the full unitary $U_{\mathrm{prep}}^\dagger \,e^{-i{H}T_G}\,U_{\mathrm{prep}}$ to $\ket{\psi_0}$, measuring in the original basis, and counting the number of times the state $\ket{\psi_0}$ is recovered (as described above).

\subsection{State preparation and un-preparation using the quantum adiabatic algorithm}

The main experimental difficulty of the compute-uncompute method outlined above to compute overlaps in general, and the Loschmidt echo in particular, is the implementation of $U_{\mathrm{prep}}^{\dagger}$ [see Eq.~\eqref{eq:echo_SM_uprep}]. Typically, states in analog quantum simulators are prepared using the quantum adiabatic algorithm (QAA). This state preparation method relies on the adiabatic theorem, which states that an initial eigenstate $\ket{\psi_0}$ of a paramete--dependent Hamiltonian ${H}(s)$ remains in the instantaneous eigenstate if the Hamiltonian is varied sufficiently slowly and the eigenstate remains separated from the rest of the spectrum by a non-vanishing gap~\cite{Albash2018}.

The QAA considers an initial Hamiltonian ${H}_0$ and a final Hamiltonian ${H}_1$, interpolating between them as
\begin{equation}\label{eq:SM_H(s)}
{H} (s) = \left[1-\gamma(s)\right]{H}_{0}+\gamma(s){H}_{1}\,,
\end{equation}
with $\gamma(s)$ a function such that $\gamma(0)=0$ and $\gamma(1)=1$. The parameter $s=t/T$ rescales physical time $t$ to the total evolution time $T$. Applying this algorithm, the state-preparation unitary $U_{\mathrm{prep}}$ is expressed as
\begin{equation}~\label{eq:SM_Uprep}
U_{\mathrm{prep}} = \mathcal{T}_{\leftarrow}\left[e^{-i\,T\int_{0}^1 {H}\left(s\right)\mathrm{d}s}\right]\,,
\end{equation}
where $\mathcal{T}_{\leftarrow}[\mathrm{exp}(\cdot)]$ denotes the time-ordered exponential, with operators ordered from right to left.

The QAA is particularly suitable for analog quantum simulation because it only requires global control of the Hamiltonian, a capability typically available in these platforms. Furthermore, it provides rigorous bounds on the runtime. An upper bound for the adiabatic time $T_\mathrm{a}$ required to prepare a state $\ket{\psi}$ with fidelity $|\langle\psi|\varphi_{0}\rangle|^2 = 1-\epsilon$ scales as
\begin{equation}~\label{eq:SM_Ta}
T_{\mathrm{a}} = \mathcal{O}\left(\max_s\frac{\lVert\partial_s H(s)\rVert+\lVert\partial^2_s H(s)\rVert}{\epsilon^{-1/2}\Delta^2(s)}+\frac{\lVert\partial_s H(s)\rVert^2}{\epsilon^{-1/2}\Delta^3(s)}\right)\,,
\end{equation}
where $\lVert\cdot\rVert$ denotes the Hamiltonian norm~\cite{Albash2018,Jansen2007,Amin2009,irmejs2025quasiadiabaticprocessingthermalstates}. Equation~\eqref{eq:SM_Ta} shows that the time dependence of the adiabatic error can be controlled via $\gamma(s)$: for the standard linear ramp $\gamma(s)=s$ (the one that we use in the main text), the error scales as $\epsilon = \mathcal{O}\left(1/T_{\mathrm{a}}^2\right)$. If sufficiently many derivatives of $\gamma(s)$ vanish at $s=0$ and $s=1$, the error can scale as $\mathcal{O}\left(e^{-cT_{\mathrm{a}}}\right)$, with $c$ a constant.

Equation~\eqref{eq:SM_Uprep} also provides a natural way to implement $U_{\mathrm{prep}}^\dagger$, by reversing the evolution and flipping the sign of the Hamiltonian, ${H}(s)\rightarrow-{H}(s)$, evolving from $s=1$ to $s=0$:
\begin{equation}~\label{eq:SM_Uprepdag}
U_{\mathrm{prep}}^\dagger = \mathcal{T}_{\rightarrow}\left[e^{+i\,T\int_{0}^1 {H}\left(s\right)\mathrm{d}s}\right]\,,
\end{equation}
where $\mathcal{T}_{\rightarrow}[\mathrm{exp}(\cdot)]$ is the reversed time-ordered exponential~\cite{AltlandSimons2010}. Flipping the sign of the integrand in Eq.~\eqref{eq:SM_Uprepdag} is crucial to obtain a proper $U_{\mathrm{prep}}^\dagger$ that can reverse the effect of $U_{\mathrm{prep}}$ in general. If only the inverse time ordering is applied, then $U_{\mathrm{prep}}^\dagger$ (up to the adiabatic error) correctly reverses $U_{\mathrm{prep}}$ only when the initial state before its application is an eigenstate of ${H}_1$. For a general state $\ket{\psi} = \sum_n c_n \ket{\varphi_n}$, reversing the adiabatic path alone does not, in general, invert the state preparation, because $\ket{\psi}$ has finite support on excited states that may experience energy level crossings during the adiabatic sweep, even in the presence of a gap.

\subsection{Measurement protocol for the Ising and the Fermi-Hubbard models}

The change of sign in ${H}(s)$ is relatively straightforward in superconducting circuits and trapped ions, as has been demonstrated experimentally in Refs.~\cite{Havlicek2019,Mi2021,Braumuller2022,abanin2025constructiveinterferenceedgequantum} and~\cite{Garttner2017,Gilmore2021,Hemery2024}, respectively. For atomic platforms, however, implementing time-reversed dynamics can be more challenging. Both the Ising model for Rydberg atoms in tweezer arrays
\begin{equation}~\label{eq:H_Ising_SM}
{H}_{\mathrm{TFIM}} = -J \sum_{\langle \bm{i}, \bm{j} \rangle} \spin{i}{x} \spin{j}{x} - g \sum_{\bm{i}} \spin{i}{z}\,,
\end{equation}
where we consider for simplicity an approximation of the Rydberg interaction as nearest-neighbor, and the Fermi-Hubbard (FH) Hamiltonian for ultracold atoms in optical lattices
\begin{equation}
{H}_{\mathrm{FH}}=-t\sum_{\langle\ii,\jj\rangle,\sigma} {c}_{\ii,\sigma}^\dagger {c}^{\vphantom{\dagger}}_{\jj,\sigma}+U \sum_{\ii}{n}_{\ii,\uparrow}{n}_{\ii,\downarrow}\,,\label{eq:H_FH_SM}
\end{equation}
have a similar structure: a term connecting distinct lattice sites (interaction for Rydberg atoms, tunneling for the Hubbard model) and a term acting locally on individual sites (transverse field or on-site interaction). Reversing the sign of the latter is experimentally simpler: in Eq.~\eqref{eq:H_Ising_SM}, $g\rightarrow -g$ can be achieved by changing the phase of the Rabi drive controlling the two-photon transition between the atomic ground and Rydberg states~\cite{Morgado2021}, whereas in Eq.~\eqref{eq:H_FH_SM}, $U\rightarrow -U$ can be realized by tuning the scattering length near a Feshbach resonance~\cite{Bloch2008Review}. Reversing the site-to-site terms, however, is more challenging.

For bipartite 2D lattices such as the square lattices considered in this work, this difficulty can be circumvented via a gauge transformation using the unitary
\begin{equation}
U_{\pi/2} = e^{i\frac{\pi}{2} {H}_{\mathrm{stagg}}}\,,
\end{equation}
with ${H}_{\mathrm{stagg}} = \sum_{\bm{i}} (-1)^{i_x + i_y}\, {n}_{\bm{i}}$ for the FH model and ${H}_{\mathrm{stagg}} = \sum_{\bm{i}} \frac{1}{2}\left[1-(-1)^{i_x + i_y}\right] \, {\sigma}_{\bm{i}}^z$ for the Ising model. This transformation flips the sign of the tunneling and Rydberg interactions, since $U_{\pi/2}^\dagger\,{c}^\dagger_{\bm{i}}{c}_{\bm{j}}\,U_{\pi/2} = e^{i\frac{\pi}{2}\left[(-1)^{i_x + i_y}-(-1)^{j_x + j_y}\right]}{c}^\dagger_{\bm{i}}{c}_{\bm{j}} = -{c}^\dagger_{\bm{i}}{c}_{\bm{j}}$, and analogously for $\spin{i}{x}\spin{j}{x}$. Importantly, this gauge transformation does not affect single-site terms. As a result, $U_{\mathrm{prep}}^\dagger$ can be implemented as
\begin{equation}~\label{eq:SM_UPrep_staggered}
U_{\mathrm{prep}}^\dagger = e^{i\frac{\pi}{2} {H}_{\mathrm{stagg}}}\,{\mathcal{T}}_{\rightarrow}\left[e^{-i\,T\int_{0}^1 {H}_{\mathrm{flip}}\left(s\right)\mathrm{d}s}\right]e^{-i\frac{\pi}{2} {H}_{\mathrm{stagg}}}\,,
\end{equation}
where ${H}_{\mathrm{flip}}$ is the Hamiltonian ${H}$ with only the single-site terms inverted in sign, i.e., if ${H} = \sum_{\langle\bm{i},\bm{j}\rangle}{H}_{\bm{i},\bm{j}}+\sum_{\bm{k}}{H}_{\bm{k}}$, then ${H}_{\mathrm{flip}}=\sum_{\langle\bm{i},\bm{j}\rangle}{H}_{\bm{i},\bm{j}}-\sum_{\bm{k}}{H}_{\bm{k}}$, with $\ii$, $\jj$, and $\kk$ running over lattice sites.

\begin{figure}[t]
    \centering
    \includegraphics[width=0.7\linewidth]{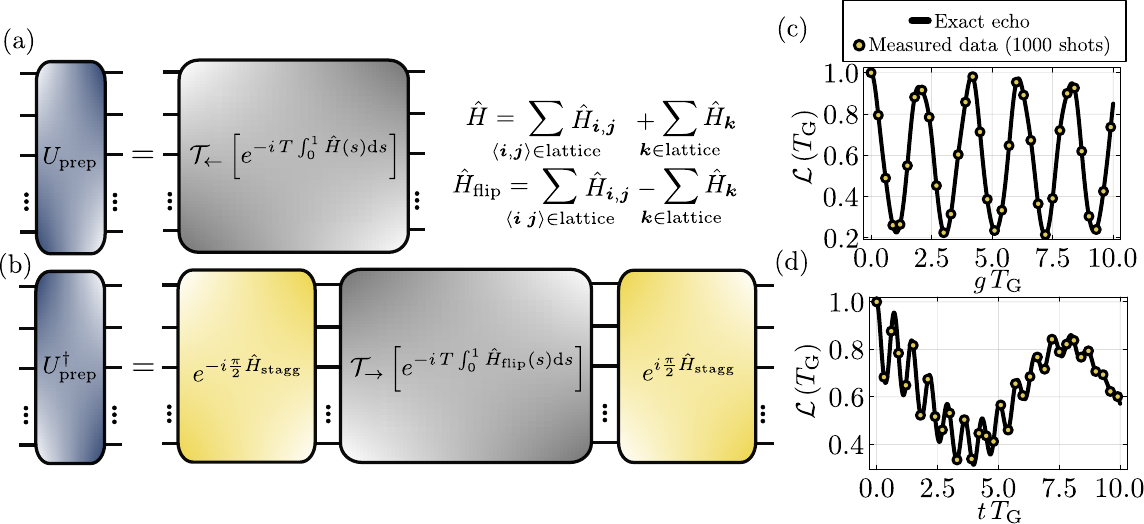}
    \caption{(a) Circuit implementing the unitary $U_{\mathrm{prep}}$ considered in the manuscript, as an adiabatic evolution ${H} (s) = \left[1-\gamma(s)\right]{H}_{0}+\gamma(s){H}_{1}$ in total time $T$. We denote the time-ordered exponential as $\mathcal{T}_{\leftarrow}$. (b) Circuit for the implementation of $U_{\mathrm{prep}}^\dagger$ for the type of Hamiltonians considered in the manuscript. The staggered rotation under ${H}_{\mathrm{stagg}}$ flips the sign of product of operators acting on two disjoint lattice sites while keeping the single-site operators invariant. This means that an effective evolution under $-{H}(s)$ can be obtained if the system is then evolved under ${H}_{\mathrm{flip}}$, where only the single-site operators are flipped. Furthermore, we need to evolve in time using the reversed time-ordered exponential, $\mathcal{T}_{\rightarrow}$. (c-d) Exact and measured echoes after simulating 1000 shots per data point for $N=4$ plaquettes of the Ising [(c)] and FH [(d)] Hamiltonians, applying $U_{\mathrm{prep}}^\dagger$ using the circuit in (b) and measuring the echoes as described in Eq.~\eqref{eq:echo_SM_uprep}. The states are adiabatically prepared as described in the text, with times $g\,T_\mathrm{a}=2$ for the Ising model and $t\,T_\mathrm{a}=12$ for the FH.} 
    \label{fig:SMUprep}
\end{figure}

In Fig.~\ref{fig:SMUprep}(a), we show the circuit implementing $U_\mathrm{prep}$ using the QAA over a total time $T$, while Fig.~\ref{fig:SMUprep}(b) illustrates the implementation of $U_\mathrm{prep}^\dagger$ according to Eq.~\eqref{eq:SM_UPrep_staggered}. The staggered rotations are applied before and after the reversed adiabatic evolution to flip the sign of the two-site terms. Figures~\ref{fig:SMUprep}(c) and (d) demonstrate the echo measurement using Eq.~\eqref{eq:echo_SM_uprep} for Ising and FH plaquettes, respectively. In Fig.~\ref{fig:SMUprep}(c), the initial state is a ferromagnetic product state with $J$ ramped adiabatically over $g\,T_\mathrm{a}=2$; in Fig.~\ref{fig:SMUprep}(d), the initial state has double occupation on two contiguous sites and empty states on the other two, with tunneling and chemical potential ramps following Ref.~\cite{Trebst2006} over $t\,T_{\mathrm{a}}=12$. Shot noise is simulated using $M=1000$ measurements per data point. In both cases, the agreement with exact echoes is excellent, showing that flipping only single-site terms is sufficient to measure the echo of an adiabatically prepared state.

Finally, we note that this approach is not limited to adiabatic state preparation but applies to general quantum circuits constructed from Hamiltonian evolutions. Variational methods~\cite{Kokail_2019,Tabares_2023,Tabares2025} may further reduce implementation time, making them attractive for experiments. Moreover, this technique has broader applications in metrology~\cite{Colombo2022} and quantum phase estimation~\cite{Yang2024PhaseMeas,Schiffer_2025,cavallar2025phasesensitivemeasurementsfermihubbardquantum}.

As a technical remark, the time-ordered exponentials in Eqs.~\eqref{eq:SM_Uprep} and~\eqref{eq:SM_Uprepdag} are implemented numerically via a finite-step Trotterization. Throughout this work, we use a second-order Trotter decomposition~\cite{Suzuki1976} for the adiabatic evolution:
\begin{equation}~\label{eq:Uprep_trotter}
U_{\mathrm{prep}} = \prod_k e^{-i{H}_0\Delta T/2}\,e^{-i{H}_1 \Delta T_k}\,e^{-i{H}_0\Delta T/2}+\mathcal{O}\left(T_\mathrm{a}\Delta T^2\right)\,,
\end{equation}
with $\Delta T_k = \frac{(T_k + T_{k+1})\Delta T}{2 T_{\mathrm{a}}}$, total preparation time $T_\mathrm{a}$, and timestep $\Delta T$. In all simulations, we take $\Delta T = 0.1$ in the appropriate units.

\section{Advantages of compressed sensing for the echo processing~\label{secSM:CompressedSensingAndNumberOfMeasurements}}

In this section, we explain why we adopt compressed sensing (CS)~\cite{Candes2006a,Candes2006b,Candes2006c} as the signal processing framework throughout the manuscript. We begin with a brief introduction to CS in Sec.~\ref{subsec:intro_CS}. Then, focusing on a specific example of echo measurements, Sec.~\ref{subsec:sampling_adv} numerically illustrates the advantages of CS compared to other methods.

\subsection{General remarks about compressed sensing}~\label{subsec:intro_CS}

CS is a signal processing technique to efficiently reconstruct a signal solving an undetermined linear system. Its most relevant advantage against other methods is its capability to recover a signal of length $N$ made of a combination of $K$ oscillating terms from \emph{very few} measurements $M$, of the order of $M=\mathcal{O}\left(K\log N\right)$. This result is achieved assuming two properties: first, the sparsity of the signal, that is, the fact that $K\ll N$ (which is a reasonable condition for real-world data); and second, that the basis of oscillatory terms (which can be cosines, exponentials or other functions) fulfill certain properties discussed below. Reducing the number of samples needed to reconstruct the signal is of significant interest in analog quantum simulators, specially those based on AMO platforms, since their small repetition rate makes the accumulation of large amounts of data challenging~\cite{Chalopin_2025,su2025fast}. 

In what follows, let $y(t)$ be a general, time dependent signal, corresponding to the Loschmidt echo in our case. This signal may be sampled at certain times $t_n$, and we denote this set as $\mathcal{T}=\left\{t_n\right\}_{i=0}^{N-1}$. The signal is then an $N-$dimensional vector, $y\in\mathbb{R}^N$, and each of its entries can be expressed in an orthonormal basis $C$ as:
\begin{equation}~\label{eq:SM_periodic_function}
    y(t_n) = \sum_{k=1}^N x_k\, c_{k} (t_n)\,.
\end{equation}
In the equation above, $x_k$ are the signal coefficients (which can be interpreted as its Fourier amplitudes, that is, its components in frequency space) and the different $c_{k} (t_n)$ are the functions spanning the orthonormal basis evaluated at time $t_n$. For instance, for the Loschmidt echo considered in the manuscript, $\mathcal{L}(T_\text{G}) = \mathcal{L}_0 + \sum_{n< m} 2\, p_n p_m \cos{\left[(E_n - E_m) T_\text{G}\right]}$, $x_k = 2\, p_n p_m \sqrt{N}$ and $c_k (t_n) = 1/\sqrt{N}\cos{\left[(E_n - E_m) t_n\right]}$ (where we have introduced the normalization factors that make the cosines an orthonormal basis for time-periodic signals). Furthermore, Eq.~\eqref{eq:SM_periodic_function} suggests that the complete vector $y$ can be written in matrix notation. Introducing the orthonormal matrix $C_{nk} = c_k(t_n)$, that is, the matrix with the waveforms $c_k$ for each value of $t_n$ as columns, then $y = Cx$, or equivalently, $x=C^\dagger y$. In practice, this means that knowing $C$ allows the computation of the non-zero frequencies of the signal and their corresponding amplitudes through this last equality. This is the spirit underlying the fast Fourier transform (FFT) algorithm and the simplest way to recover the properties of the signal.

In the limit of $N\rightarrow\infty$, the sum in Eq.~\eqref{eq:SM_periodic_function} contains infinitely many frequencies and can therefore approximate any signal. For finite $N$, however, not all the signal frequencies may be contained in the possible values for the frequencies. For instance, for the discrete cosine transform~\cite{BritanakYipRao2006DCTDST} of interest in our manuscript (the transform that expresses time-dependent signals as linear combination of cosines), the available frequencies are $\mathcal{W} = \left\{\pi k/N\right\}_{k=0}^{N-1}$. If $k\in\mathbb{R}$, then all the possible frequencies are contained in the set $\mathcal{W}$ (which is itself $\mathbb{R}$); however, since there are $N$ of these values, the maximum resolution in frequency space that we should expect from the application of this method is $\sim 1/N$. In practice this means that, if at least one of the signal true frequencies lies outside this set, $\omega_k\notin\mathcal{W}$, which is referred as an off-grid frequency, then computing the signal properties as $x=C^\dagger y$ (or equivalent, with the FFT) leads to spectral leakage; that is, a blurring of the actual values of the frequencies and its amplitudes. This ultimately means that the inaccuracy recovering the signal parameters will decrease as $1/N$, requiring a huge amount of samples to accurately estimate them.

However, as long as the signal is \emph{sparse} (that is, that the number of different oscillating terms $K$ in Eq.~\eqref{eq:SM_periodic_function} is such as $K\ll N$), most of the columns in the matrix $C$ will not play a role in the computation $x=C^\dagger y$. Moreover, if the matrix $C$ fulfills a \emph{restricted isometry property} (RIP)~\cite{Candes2008_RIP} (which means that one can define an effectively approximate orthogonal matrix when sampling its rows or columns randomly), then the relevant entries of $C$ would still define an effective discrete cosine transform. Inspired by this, in Ref.~\cite{Candes2006a} the authors realized that one could almost always recover $x$ exactly from $y$ by solving the following convex program:
\begin{equation}~\label{eq:appendix_CS_program_noiseless}
    \min_{x\in\mathbb{R}^N}\lVert x\rVert_1 \quad\mathrm{s.t.}\quad  C_{\Omega}^\dagger x = y_{\Omega}\,,
\end{equation}
where $\lVert\cdot\rVert_p$ denotes the $p$-norm of the vector (for a vector $x$, this means that $\lVert x \rVert_p = \left(\sum_i |x_i|^p\right)^{1/p}$) and $\Omega \subset \mathcal{T}$ is a randomly chosen subset of times (so $y_{\Omega}$ and $C_{\Omega}^\dagger$ denote the projections of $y$ and $C$ over $\Omega$). 

A remarkable result is that the size of $\Omega$ needs to scale only logarithmically with the desired resolution of the recovered signal, that is, $|\Omega|=\mathcal{O}\left(\log{N}\right)$. In practical experiments, $\Omega$ can be precomputed so the data is only sampled at those $t_n$. The accuracy of the components, however, can be greatly improved, since the only theoretical limit for the size of the set of frequencies $\mathcal{W}$ are the (classical) computational resources needed to solve the convex program in Eq.~\eqref{eq:appendix_CS_program_noiseless}. In practice, however, these frequencies cannot be continuous unless atomic norm methods are used~\cite{candes2012mathematicaltheorysuperresolution,candes2013superresolutionnoisydata,chisuper2020} (which are beyond the scope of this paper). This means that the number of total frequencies is also bounded, and thus some degree of spectral leakage will also be observed. However, this will be exponentially smaller than the corresponding leakage for the equivalent discrete cosine transform applied to the same number of measured data points. Furthermore, if the measured signal includes some sort of noise, so $y=y^0 + z$ with $y_0$ being the noiseless oscillating signal and $z$ the noise vector, then the recovery is still possible~\cite{Candes2006b} solving
\begin{equation}~\label{eq:appendix_CS_program_noisy}
    \min_{x\in\mathbb{R}^N}\lVert x\rVert_1 \quad\mathrm{s.t.}\quad \lVert C_{\Omega}^\dagger x - y_{\Omega}\rVert_2\leq\eta
\end{equation}
instead, where $\eta$ is the size of the error term $z$. As it is discussed in the main text, the error $z$ has three different origins when applying this framework to the estimation of Loschmidt echoes obtained from an experiment: (i) shot noise, arising from finite measurements; (ii) noise arising from off-grid frequencies; and (iii) noise arising from considering a number of relevant frequencies smaller than the total number of frequencies in the signal. Although CS is expected to do not suffer significantly because of this noise, to make our estimation of the signal parameters even more resilient to it, we also include in the main text a final nonlinear fit to a linear combination of cosines using as initial parameters the ones obtained from CS. This significantly reduces the effects of noise of types (i) and (ii), and as long as the non-considered frequencies are sufficiently small, CS will deal with the noise of type (iii)~\cite{Candes2007CompressiveSampling}.

\subsection{An example of the sampling advantage of compressed sensing}~\label{subsec:sampling_adv}
\begin{figure}[t]
    \centering
    \includegraphics[width=0.5\linewidth]{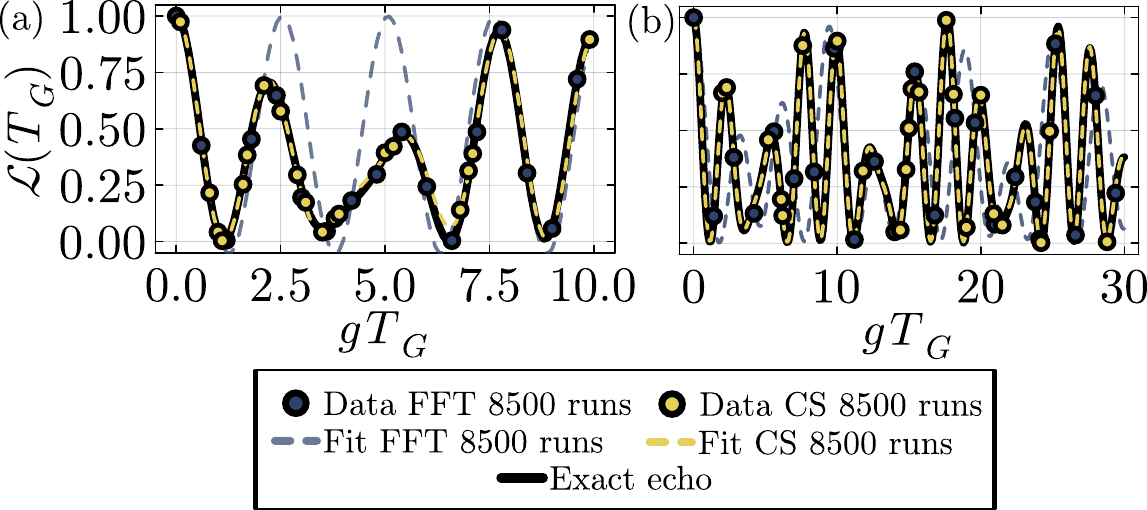}
    \caption{(a-b) Loschmidt echoes of a $2\times4$ antiferromagnetic Ising ladder with $J/g=1$ for different total times. The initial states are defined in the main text. In the CS case, we randomly choose a set of $M_1$ different times $t_n$ [$17$ for (a) and $23$ for (b)] and simulate approximately $M_2\sim500$ measurements for each $t_n$, so the total number of runs is $M=M_1\times M_2 = 8500$. We then use the computed data to implement the noisy version of the algorithm, solving the problem in Eq.~\eqref{eq:appendix_CS_program_noisy} according to the details in the End Matter (including the nonlinear fit). For FFT, we equally-space the same number of $M_1$ points and estimate the data for $M_2\sim500$ measurements too, so $M$ is the same in both cases. We follow the procedure outlined in the text to obtain the final fit to the signal.}
    \label{fig:SM_CS1}
\end{figure}

We now check the advantage of compressed sensing against a standard FFT in terms of sampling cost for the specific case of estimating the properties of a Loschmidt echo. We define this sampling cost as the total number of experimental runs $M$ needed to acquire the echo data. Typically, $M=M_1\times M_2$ is determined by the $M_1$ different values of $t_n$ at which data is acquired, and also by the $M_2$ total shots that need to be performed at each value of $t_n$ to estimate the echo with an error $\mathcal{O}(1/\sqrt{M_2})$. 

For the examples in Fig.~\ref{fig:SM_CS1}, we initialize a $2\times4$ Ising ladder with $J/g=1$ in the state $\ket{\psi} = \sqrt{0.45}\ket{\varphi_0}+\sqrt{0.35}\ket{\varphi_3}+e^{-i\frac{\pi}{2}}\sqrt{0.2}\ket{\varphi_4}$, $\left\{\ket{\varphi_n}\right\}$ being the eigenstates of the Ising Hamiltonian ${H}_{\text{TFIM}}$ for the ladder. Then, we evolve the system under ${H}_{\text{TFIM}}$ for a total time $T_{\mathrm{G}}$ [$g\,T_{\mathrm{G}}=10$ in the case of Fig.~\ref{fig:SM_CS1}(a) and $g\,T_{\mathrm{G}}=20$ for Fig.~\ref{fig:SM_CS1}(b)] and measure the corresponding echo for different values of $t_n$. We consider $M_1=17$ and $M_1=19$ different times for Figs.~\ref{fig:SM_CS1}(a) and (b), respectively, and adjust the number of measurements $M_2$ for each case so both would require the same number of experimental runs $M=M_1\times M_2 =8500$ to obtain the data. For the CS case, the $M_1$ data points are randomly chosen; for the FFT, however, they are equally spaced. In the FFT case, moreover, we first apply to the signal a Hann window to suppress spectral leakage~\cite{OppenheimSchaferBuck1999DTSP2e}, followed by the FFT, a sub-bin refinement using a parabolic interpolation to fix the amplitudes and then a final least-squares fit to the amplitudes. As we do for the CS case, these values are then fed as the initial parameters of a nonlinear regression to refine their values.

The results in both Figs.~\ref{fig:SM_CS1}(a) and~(b) are clear: the fit that we obtain using the CS procedure accurately recovers the echo signal, showing great agreement with the exact value of the echo. The FFT based procedure, nevertheless, offers a very poor approximation if the total time available is $g\,T_{\mathrm{G}}=10$, as it is the case of Fig.~\ref{fig:SM_CS1}(a), but improves its accuracy when $g\,T_{\mathrm{G}}=20$ in Fig.~\ref{fig:SM_CS1}(b). This is expected, since the frequency resolution of the FFT is approximately given by $1/(g\,T_{\mathrm{G}})$, so doubling it reduces spectral leakage and gives good initial estimates for the final nonlinear fit. However, none of these results nearly match the accuracy obtained with the CS procedure when the number of experimental runs in both cases is the same. This highlights the CS procedure as doubly advantageous for sampling data from current analog quantum simulators, since it needs less evolution time $T_{\mathrm{G}}$ (therefore minimizing the effects of decoherence) while also requiring less total samples (so it can be practically implemented with current repetition rates).

\begin{figure}[t]
    \centering
    \includegraphics[width=0.5\linewidth]{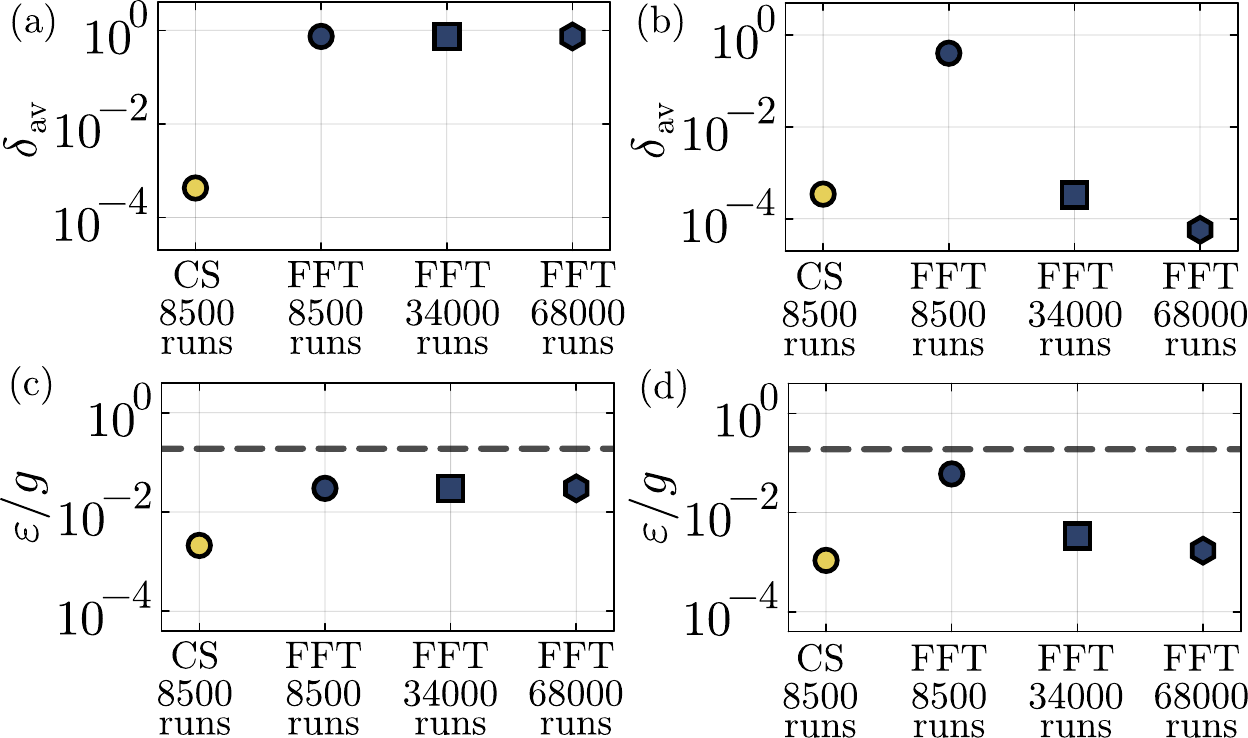}
    \caption{(a-b) Average error in the estimation of the signal parameters using the echo for the CS and FFT based procedures considering a total number of experimental runs $M=M_1\times M_2$. The total time of the measured signal is $g\,T_{\mathrm{G}}=10$ in (a) and $g\,T_{\mathrm{G}}=20$ (b). The results corresponding to 8500 runs consider $M_1 = 17$ in (a) and $M_2=500$ in (b), adjusting the values of $M_2\sim 500$ accordingly. For the rest of the FFT cases, more (equally spaced) $M_1$ sampling points are obtained while requiring $M_2\sim 500$. (c-d) Residual energies after applying the GENTLE protocol of the main text using the results of the fits in (a) and (b) for (c) and (d), respectively. The initial value of the residual energy is shown as a black dashed line.}
    \label{fig:SM_CS2}
\end{figure}

To further quantify the accuracy of the different methods, we introduce the average error in the estimation of the signal parameters as:
\begin{equation}
    \delta_{\mathrm{av}} = \frac{1}{2P}\sum_{p=1}^P\left(\left|\omega^{\mathrm{ex}}_p - \omega^{\mathrm{fit}}_p\right|+\left|A^{\mathrm{ex}}_p - A_p^{\mathrm{fit}}\right|\right)\,,
\end{equation}
where the $\omega^{\mathrm{ex}}_p$ and $A^{\mathrm{ex}}_p$ are, respectively, the $P$ exact frequencies and amplitudes of the signal, while $\omega^{\mathrm{fit}}_p$ and $A^{\mathrm{fit}}_p$ are the ones obtained from the fit (some of the latter may be zero). In Figs.~\ref{fig:SM_CS2}(a) and~(b) we show $\delta_{\mathrm{av}}$ for data obtained at $g\,T_{\mathrm{G}} = 10$ and $g\,T_{\mathrm{G}} = 20$, respectively. For the first case, Figs.~\ref{fig:SM_CS2}(a), the CS procedure is able to perfectly recover the signal parameters, but the FFT cannot. This does not change even if we multiply the sampling rate by $4$ or even by $8$, since the main limitation in this case is the limited frequency resolution that the FFT has for the small value of $g\,T_{\mathrm{G}}=10$ considered. This situation changes in Figs.~\ref{fig:SM_CS2}(b), where now the results obtained from the FFT match the ones with CS once the number of samples $M_1$ is multiplied by $4$ with respect to the CS case. This is, first, because in this case we are considering a total signal time of $g\,T_{\mathrm{G}}=20$, doubling the frequency resolution with respect to Figs.~\ref{fig:SM_CS2}(a). But the fact that we need more time samples to achieve the accurate results is also because this procedure is more sensitive to noise than the CS analysis, so it needs more data points to oversample and reduce the effects of the noisy measurements. 

Finally, in Figs.~\ref{fig:SM_CS2}(c) and~(d) we show the results of applying the GENTLE protocol to estimate the ground state energy using as inputs the signal parameters from Figs.~\ref{fig:SM_CS2}(a) and~(b), respectively. As expected, in both cases the CS procedure is able to increase the energy estimation by almost two orders of magnitude requiring much less evolution time $T_{\mathrm{G}}$ and also much less total experimental runs $M$. The parameters coming from the FFT procedure, nevertheless, need both more $T_{\mathrm{G}}$ and more $M$ to yield similar results to the CS for the ground state energy.  Finally, let us note that in the manuscript we do not randomize the times, but instead apply the CS algorithm directly over equally-spaced $t_n$ measurements. This does not fully exploit the advantages of CS, but as the results in the main text show, it still demonstrates an advantage. In this section, however, we have demonstrated with a numerical example that such advantage holds when the number of $t_n$ chosen is exponentially smaller, as long as these are randomly selected.

\section{Analysis of finite temperature effects~\label{secSM:ExperimentalAnalysis}}

The discussion in the main text considers pure product states $\ket{\psi_0}$ as the starting point of the protocol. These states are then transformed, using the unitary $U_{\mathrm{prep}}$ discussed in Sec.~\ref{secSM:StatePreparation}, into a state $\ket{\psi}$ that has enough overlap with the ground state. This state is then evolved under the Hamiltonian ${H}$ and its Loschmidt echo, $\mathcal{L}(T_{\text{G}}) = |\left\langle\psi\right|e^{-i{H}T_{\text{G}}}\left|\psi\right\rangle|^2$, is measured. However, analog quantum simulators do not always prepare perfect pure states, but mixed states due to finite-temperature effects. This is specially significant in the case of ultracold atoms in optical lattices, where state preparation methods start from a non-zero entropy state whose Hamiltonian is adiabatically deformed~\cite{Bloch2008Review}. Although the entropy of the initial state can be very low, using for instance Mott~\cite{Dai2016,Yang2020_cooling,Yang2020GaugeInvariance,Zhang2023_atoms} or band~\cite{chiu2018,yan2022,spar2022,Xu2025_nature} insulators product states that are adiabatically connected to the target many-body state~\cite{Paredes2008,Rey2009,Lubasch2011,Nascimbene2012}, the initial states are still at a finite temperature. 

The adiabatic theorem is formulated for pure states, so a complete understanding of when and why the quantum adiabatic algorithm succeeds preparing a target state when the initial state is thermal is still an open question~\cite{Carcy2021CertifyingAdiabaticBosons,Zuo2024WorkStatisticsAdiabatic,Greenblatt2024AdiabaticLowTemp,irmejs2025quasiadiabaticprocessingthermalstates}. For the same reason, a natural follow-up question is whether it is possible to measure an object equivalent to the Loschmidt echo in Eq.~\eqref{eq:SM_Uprep} when finite-temperature effects are introduced, and whether such object can be later used to estimate the spectral information needed to apply our algorithm.

This section demonstrates that the GENTLE protocol also works when considering initial thermal states. First, in Sec.~\ref{subsec:SM_thermal_analytic} we compute the Loschmidt echo in terms of density matrices for an initial Gibbs state, and derive the conditions of validity for the recovery of the information needed for the protocol. Then, in Sec.~\ref{subsec:SM_thermal_numeric} we numerically check this result for small instances of the FH and the Ising models. This benchmark demonstrates that the spectral information of the system can be recovered from the echo measured over an initial thermal state, provided that the temperature is small enough, and that such information is still an accurate input to solve the nonlinear equations of the GENTLE protocol.

\subsection{The GENTLE protocol for finite-temperature initial states~\label{subsec:SM_thermal_analytic}}

Let us consider an easy to prepare Gibbs state $\rho_0$ at inverse temperature $\beta$, corresponding to a Hamiltonian ${H}_0$ whose ground state is a product state, 
\begin{equation}~\label{eq:SM_rho0_thermal}
    \rho_0 = \frac{e^{-\beta{H}_0}}{\mathcal{Z}_0}=\frac{1}{\mathcal{Z}_0}\sum_k e^{-\beta E_{k}^{0}}\left|\varphi_k^ 0\right\rangle\left\langle\varphi_k^0\right|\,,\quad\text{with}\quad\mathcal{Z}_0=\tr\left[e^{-\beta{H_0}}\right],
\end{equation}
where $\left\{\left|\varphi_k^0\right\rangle\right\}$ and $\left\{E_k^0\right\}$ are the eigenstates and eigenenergies of ${H}_0$, respectively. The initial state for the measurement of the echo is then prepared by applying the same unitary $U_\text{prep}$ as discussed in the main text, $\rho = U_\text{prep}\rho_0 U_\text{prep}^\dagger$.  In the zero temperature case, the corresponding initial state  $\ket{\psi}$ can be expanded in the basis of eigenstates of the target Hamiltonian ${H} = \sum_n E_n\left|\varphi_n\right\rangle\left\langle\varphi_n\right|$ as $\ket{\psi}=\sum_n c_n \ket{\varphi_n}$, with a corresponding density matrix $\rho_{\psi} = \ket{\psi}\bra{\psi}=\sum_{n,m} c_n c_m^*\ket{\varphi_n}\bra{\varphi_m}$. In the finite temperature case, however, the density matrix $\rho$ is expressed as:
\begin{equation}
    \rho = U_\text{prep}\,\rho_0\,U_\text{prep}^\dagger= Z_\beta\sum_{n,m}c_n c_m^*\ket{\varphi_n}\bra{\varphi_m}+\left(1-Z_\beta\right)\sum_{n,m}a_n a_m^*\ket{\varphi_n}\bra{\varphi_m}=\sum_{n,m}\rho_{nm}\ket{\varphi_n}\bra{\varphi_m}\,,
\end{equation}
with $Z_\beta = e^{-\beta E_0^0}/\mathcal{Z}_0$. In this expression, we explicitly express each coefficient $\rho_{nm}$ as a combination of its zero-temperature version and the terms arising from the finite temperature contributions, parametrized with the coefficients $\left\{a_n\right\}$ arising from the action of $U_\text{prep}$ over the excited eigenstates. 

The state $\rho$ is then evolved under the Hamiltonian ${H}$, leading to $\rho\left(T_{\mathrm{G}}\right)=e^{-i T_{\text{G}}{H}}\,\rho\,e^{i T_{\text{G}}{H}}$, and then projected back to $\rho$, resulting in
\begin{equation}~\label{eq:SM_echo_density}
    \tr\left[\rho\,\rho\left(T_{\mathrm{G}}\right)\right] = \tr\left[\rho_0 \,U_\text{prep}^\dagger e^{-iT_{\text{G}}{H}}U_\text{prep}\,\rho_0 \,U_\text{prep}^\dagger e^{iT_{\text{G}}{H}} U_\text{prep}\right] = \sum_{n} \rho_{nn}^2 + 2\sum_{n<m}\left|\rho_{nm}\right|^2\cos{\left[\left(E_n - E_m\right)T_\text{G}\right]}\,,
\end{equation}
where in the second equality we have used the cyclic property of the trace.
As in the zero-temperature case, this unitary involves the application of $U_{\text{prep}}$, the time evolution under ${H}$, and the reversal operation $U_{\text{prep}}^\dagger$, as discussed in Sec.~\ref{secSM:StatePreparation}. 
Let us also note that a full generalization of the Loschmidt echo for density matrices involves using the Uhlmann fidelity and computing square roots of density matrices, and that Eq.~\eqref{eq:SM_echo_density} is instead the linearization of such quantity  (see, for instance, Ref.~\cite{Jacobson2011_echothermal} for a discussion). Nevertheless, the analytical form of Eq.~\eqref{eq:SM_echo_density} as a linear combination of oscillating cosines is exactly what we need for the protocol while still being relatively straightforward to measure, so we keep $\tr\left[\rho\,\rho\left(T_{\mathrm{G}}\right)\right]$ as the generalization of $\mathcal{L}(T_G) = |\left\langle\psi\right|e^{-i{H}T_G}\left|\psi\right\rangle|^2$.

\begin{figure}[t]
    \centering
    \includegraphics[width=0.6\linewidth]{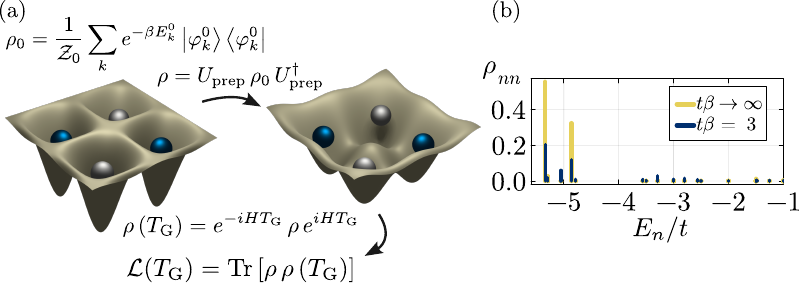}
    \caption{(a) The proposed implementation starts from an easy to prepare Gibbs state, $\rho_0$, at an inverse temperature $\beta$. In this case, the initial state is a Mott insulator product state. That state is then evolved under the unitary $U_{\mathrm{prep}}$ to prepare the initial state for the GENTLE protocol, $\rho$. Such state is evolved under the Hamiltonian $H$ at different times $T_\mathrm{G}$, and the echo is computed projecting it back to $\rho$, following the method described in \ref{secSM:StatePreparation}. (b) The LDOS of a FH plaquette [see Eq.~\eqref{eq:SM_LDOS_rho}] prepared as depicted in (a), following the adiabatic protocol discussed in Ref.~\cite{Tabares2025}, for $U/t = 1$ and $t\,T_{\mathrm{a}}=3$ at each adiabatic step. We show the zero temperature case and a finite temperature one.}
    \label{fig:SM_thermal1}
\end{figure}

To better understand the effect of a finite temperature in the measurement of the echo we can inspect each of the terms in Eq.~\eqref{eq:SM_echo_density}. First, the sum of the non-oscillating terms correspond to the purity of $\rho$ and $\rho_0$, since we assume that the state preparation and the time evolution are perfect unitaries. Since
\begin{equation}
    \rho_{nn} = Z_{\beta}\left|c_n\right|^2 + \left(1-Z_{\beta}\right)\left|a_n\right|^2\,,
\end{equation}
we see that the consequence of a finite temperature in this case is that the zero temperature contributions $|c_n|^2$ get damped by the factor $Z_\beta$, while new contributions proportional to the $|a_n|^2$ coefficients appear. There are two possible limits in this case: first, if the temperature of the initial state is low enough, then $Z_{\beta}$ can be expanded as a geometric series to yield $ Z_{\beta}\approx 1 -\sum_{k>0}e^{-\beta \left(E_{k}^{0}-E_{0}^{0}\right)}$ and $\rho_{nn} \approx |c_n|^2 - \sum_{k>0}e^{-\beta \left(E_{k}^{0}-E_{0}^{0}\right)}\left(|c_n|^2 - |a_n|^2\right)$. Typically, this means that the values of $\rho_{nn}$ corresponding to the eigenstates in which the zero temperature state has significantly overlap will decrease, while certain contributions of $\rho_{nn}$ that would be zero in the zero temperature case will be finite now (but exponentially small). The second limiting case, on the other hand, is the limit of exponentially long runtimes in the adiabatic coupling [the considerations above were valid for a general $U_{\text{prep}}$, but now we assume that this operator is obtained from Eq.~\eqref{eq:SM_Uprep}]. In such situation, and as long as there are are no degeneracies in the spectrum of ${H}(s)$ to avoid eigenstate mixing, $|c_n|^2 = \delta_{n,0}$ and $|a_n|^2 = \delta_{n,m}$, with $\delta_{n,m}$ being the Kronecker delta. Each eigenstate of ${H}_0$ in Eq.~\eqref{eq:SM_rho0_thermal} gets uniquely mapped to an eigenstate of ${H}$ and $\rho_{nn} = e^{-\beta E_{n}^{0}}/\mathcal{Z}_0$, so the contributions to each of the eigenstates will depend on the temperature of the initial state.

The two cases described above can be better understood by computing the local density of states (LDOS) for $\rho$, defined as
\begin{equation}~\label{eq:SM_LDOS_rho}
    D_{\rho}(E) = \sum_n \braket{\varphi_n|\rho|\varphi_n}\delta(E-E_n)\,.
\end{equation}
We show in Fig.~\ref{fig:SM_thermal1}(b) an explicit calculation of $D(E)$ for a FH plaquette prepared starting from a Mott insulator state which is adiabatically evolved following the protocol of Ref.~\cite{Tabares2025}. This protocol consists first in preparing adiabatically two uncoupled dimers, and then these dimers are coupled to form the plaquette. In Fig.~\ref{fig:SM_thermal1}(a) we show the LDOS for a state prepared like this with $U/t=1$ and $t\,T_{\mathrm{a}}=3$ for each of the adiabatic couplings. In the zero temperature regime and assuming a perfect adiabatic state preparation, the LDOS is simply a spike in the energy value corresponding to the ground state energy. As the adiabaticity is lost, some extra spikes appear in certain excited states, as Fig.~\ref{fig:SM_thermal1}(b) shows for $t\beta\rightarrow\infty$. If a finite 
value of $\beta$ is considered instead, such as $t\beta=3$, even before starting the adiabatic evolution the instantaneous density of states is already spread. When the adiabatic evolution is completed, as it was discussed in the analysis for $t\beta\gg1$ above, the amplitude of the spikes corresponding to the excited eigenstates (with $|c_n|^2$ contributions) relevant in the zero temperature case decrease, while new spikes corresponding to the $|a_n|^2$ terms appear. However, if $\beta$ is big enough compared to $\Delta_0 = E_{1}^{0}-E_{0}^{0} $ (the gap of ${H}_0$), the amplitude of these new spikes will be exponentially small and the LDOS at finite temperature still has as it most significant peaks in the same positions they were at zero temperature, so as long as the adiabaticity conditions required in the $\beta\rightarrow\infty$ case also hold now, the measured echo also provides spectral information in the finite temperature case. Finally, let us also comment that the procedure proposed in the main text to measure the echo implies substituting $\rho_0 \approx \left|\varphi_{0}^{0}\right\rangle\left\langle\varphi_{0}^{0}\right|$ in the final projection of Eq.~\eqref{eq:SM_echo_density}. This approximation, however, is expected to be reasonable as long as the condition $\beta\gg1/\Delta_0$ discussed above holds.

Therefore, according to Eq.~\eqref{eq:SM_echo_density}, a fit of the measured signal to a linear combination of cosines yields the frequencies $\omega_p^*=|E_n-E_m|$ and the amplitudes $A_p^* = 2|\rho_{nm}|^2$ for the oscillations. The GENTLE protocol then combines this information with the expectation values of
\begin{equation}~\label{eq:SM_expect_values}
    \langle{H}\rangle = \tr\left[\rho{H}\right] = \sum_{n}\rho_{nn}E_n\quad\text{and}\quad \langle{H}^2\rangle = \tr\left[\rho{H}^2\right] = \sum_{n}\rho_{nn}E_n^2\,.
\end{equation}
However, a key difference appears now with respect to the formulation with pure states. In such case, the amplitudes of the oscillations provide the products between coefficients $2|c_n|^2 |c_m|^2$, and each of these coefficients also appear in the expectation values $\langle {H}\rangle$ and $\langle {H}^2\rangle$. In this case, however, the coefficients appearing in the latter are the diagonal terms of $\rho$, but the terms obtained from the fit are still off-diagonal terms of the density matrix. This apparent contradiction is solved realizing that this is the case also for the density matrix of a pure state, since the relevant off-diagonal terms in such case correspond to $\left|\rho_{nm}\right|^2=2\,\left|c_n\right|^2\,\left|c_m\right|^2 = \rho_{nn} \rho_{mm}$. Hence, to apply the GENTLE protocol to an initially mixed state we need to assume that $\left|\rho_{nm}\right|^2\approx \rho_{nn} \rho_{mm}$. If that approximation holds, the protocol can be implemented without further modifications, using the amplitudes and oscillations of the measured signal as inputs to the nonlinear system of equations that also includes the expectation values in Eq.~\eqref{eq:SM_expect_values}.

We can quantify the conditions of validity of the approximation $|\rho_{nm}|^2\approx\rho_{nn}\rho_{mm}$ by independently inspecting the terms
\begin{equation}~\label{eq:SM_rhonm}
    \left|\rho_{nm}\right|^2 = Z_\beta^2 \,|c_n|^2 \,|c_m|^2 + \left(1-Z_\beta\right)^2\,|a_n|^2\,|a_m|^2+2Z_\beta(1-Z_\beta)\mathrm{Re}\left[c_n c_m^* a_n^* a_m\right]\,
\end{equation}
and
\begin{equation}~\label{eq:SM_rhonnrhomm}
     \rho_{nn}\rho_{mm} = Z_\beta^2 \,|c_n|^2 \,|c_m|^2 + \left(1-Z_\beta\right)^2\,|a_n|^2\,|a_m|^2+Z_\beta(1-Z_\beta)\left(|c_n|^2 |a_m|^2 + |c_m|^2 |a_n|^2 \right)\,.
\end{equation}
Substracting Eqs.~\eqref{eq:SM_rhonm} and~\eqref{eq:SM_rhonnrhomm} we find that:
\begin{equation}
    \Big|\left|\rho_{nm}\right|^2 - \rho_{nn}\rho_{mm}\Big| = Z_\beta(1-Z_{\beta})\Big||c_n||a_m|\big(|c_m||a_n|\cos{\phi}-|c_n||a_m|\big)+|c_m||a_n|\big(|c_n||a_m|\cos{\phi}-|c_m||a_n|\big)\Big|\,,
\end{equation}
with $\phi = \arg{\left(c_n c_m^* a_n^* a_m\right)}$. As it was discussed above, if $\beta\Delta_0\gg1$ (that is, the inverse temperature is much greater than the gap of the initial Hamiltonian), then the term $1-Z_{\beta}$ goes exponentially fast to zero with $\beta$. Furthermore, although this were not be the case, if the spectrum of ${H}(s)$ is non-degenerate, the difference $\Big||c_n||a_m|-|c_m||a_n|\Big|$ will approach zero as the adiabaticity of the state preparation is increased (even in the most common degenerate ${H}(s)$ case we expect this intuition to hold in general). Therefore, in summary, we find that as long as:
\begin{equation}~\label{eq:SM_condition_thermal}
    \beta\gg\frac{1}{E_{1}^{0}-E_{0}^{0}}\quad \text{and}\quad \braket{\varphi_0|\rho|\varphi_0}>\braket{\varphi_n|\rho|\varphi_n}\quad\text{for}\quad n>0\,,
\end{equation}
we expect the GENTLE protocol to be able to appropriate recover the energy differences and the amplitudes from the fit of the Loschmidt echo and use this information to compute the value of the ground state energy.

\subsection{Accuracy of the GENTLE protocol for finite temperatures~\label{subsec:SM_thermal_numeric}}

\begin{figure}[t]
    \centering
    \includegraphics[width=0.5\linewidth]{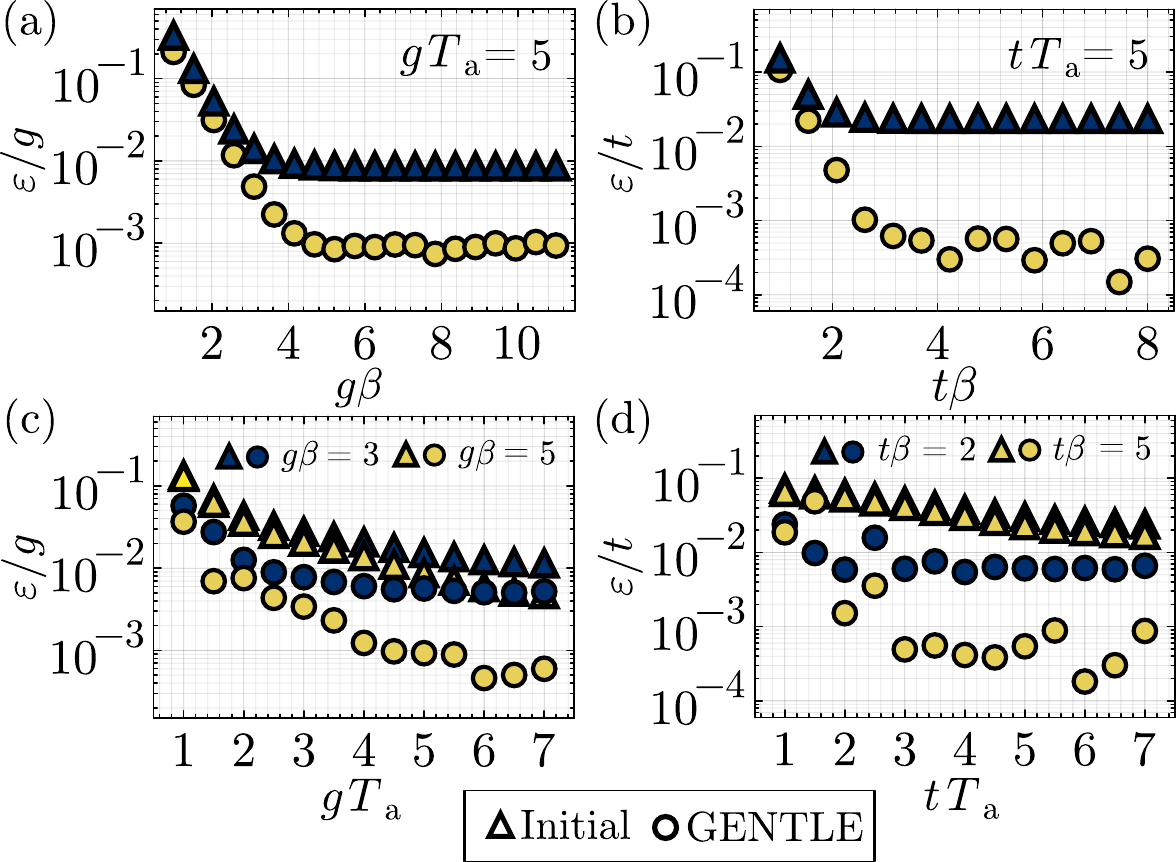}
    \caption{(a-b) Accuracy of the GENTLE protocol as a function of $\beta$ for adiabatically prepared states in a $N=8$, $g/J=0.8$ 1D Ising chain [(a)] and a $N=4$, $U/t=6$ FH plaquette [(b)], prepared in time $g\,T_{\mathrm{a}}=5$ and $t\,T_{\mathrm{a}}=5$, respectively, as described in the main text. The echo is measured for $g\,T_{\mathrm{G}}=30$ [(a)] and $t\,T_{\mathrm{G}}=34$ [(b)], considering $M=10000$ [(a)] and $M=1000$ [(b)] shots per data point and sampling logarithmically, according to Sec.~\ref{secSM:CompressedSensingAndNumberOfMeasurements} (c-d) GENTLE protocol for two fixed temperatures and different adiabatic coupling times for the Ising [(c)] and the FH [(d)] cases discussed in (a) and (b), respectively. Let us note that neither (c) or (d) apply the KDE protocol to the data, so this is the reason why there are some small fluctuations in the residual energy.}
    \label{fig:SM_thermal2}
\end{figure}

To check the accuracy of the regimes discussed above, we  check now the behavior of the GENTLE protocol when the initial states considered are thermal states. In Figs.~\ref{fig:SM_thermal2}(a) and~(b) we show, respectively, the results for a $N=8$, $g/J=0.8$ antiferromagnetic Ising chain and a $N=4$, $U/t=6$ FH plaquette. In the Ising case we start from a ferromagnetic product state, while in the FH case we prepare dimers with almost perfect fidelity [$t\,T_{\mathrm{a}}^{(\mathrm{dim})}=100$] and then couple them in a time $T_{\mathrm{a}}$. This protocol is the adapted version of the experimental protocol in Ref.~\cite{Xu2025_nature}, where dimers were prepared with almost perfect fidelity but starting from a band instead of a Mott insulator. In both cases we take $g\,T_{\mathrm{a}}=t\,T_{\mathrm{a}}=5$. As Figs.~\ref{fig:SM_thermal2}(a) and~(b) show, the accuracy of the protocol increases monotonically with $\beta$, and so do the estimation of the energy over the initial state. However, the latter saturates early, but the accuracy of the GENTLE protocol keeps improving up to saturating almost one order of magnitude later. Furthermore, $\Delta_0=2g$ in the Ising case and $\Delta_0=3t$ in the FH one; that's the reason one Fig.~\ref{fig:SM_thermal2}(a) saturates at $g\beta\approx4$ while Fig.~\ref{fig:SM_thermal2}(b) does it earlier, at $g\beta\approx2.5$. This verifies for a particular example the condition derived in the previous section [Eq.~\eqref{eq:SM_condition_thermal}], stating that the protocol works as long as $\beta\Delta_0\gg 1$.

Finally, in Figs.~\ref{fig:SM_thermal2}(c) and~(d) we do a similar analysis as the one discussed above, but now fixing two different temperatures and analyzing the accuracy as a function of the adiabaticity of the state preparation. As expected, we find a small improvement with $T_{\mathrm{a}}$ for lower values of $\beta$, demonstrating that the protocol still retains its expected zero temperature behavior (increasing the accuracy with $T_{\mathrm{a}}$) even in these cases. However, to see a significant improvement (again, of more than one order of magnitude), we need to have a value of $\beta$ such as $\beta\Delta_0\gg 1$.

\end{widetext}
\end{document}